\pdfoutput=1

\documentclass[final,twocolumn]{raa_twocolumn}            

\usepackage{graphicx,times}             
\usepackage{natbib}
\usepackage{amssymb,amsmath}
\usepackage{booktabs}
\usepackage{pdflscape}
\bibpunct{(}{)}{;}{a}{}{,}
\usepackage{soul}
\usepackage{CJK}

\usepackage{multirow}
\usepackage{makecell}
\usepackage[pagebackref=true,
            citecolor=blue,
            colorlinks=true,
            linkcolor=blue,
            filecolor=blue, 
            urlcolor=blue, 
            final=true,
            ]{hyperref}

\newcommand{\hii}{H\textsc{ii}}
\newcommand{\msun}{$ M_\odot$}

\newcommand{\kms}{km\,s$^{-1}$}
\newcommand{\jybeam}{Jy\,beam$^{-1}$}
\newcommand{\mjybeam}{mJy\,beam$^{-1}$}

\newcommand{\degree}{$^{\circ}$}
\newcommand{\parcsec}{\mbox{$.\!\!\arcsec$}}
\newcommand{\ssstyle}{\scriptscriptstyle}

\defcitealias{Csengeri2017ACA}{C17}
\defcitealias{Traficante2023SQUALO-I}{T23}

\newcommand{\htco}{H$_2$CO}

\begin{document}
\begin{CJK*}{UTF8}{gbsn}
\title{The ALMA-QUARKS Survey: II. the ACA 1.3 mm continuum source catalog and the assembly of dense gas in massive star-forming clumps}

\volnopage{Vol.0 (20xx) No.0, 000--000} 
\setcounter{page}{1} 

\author{Fengwei Xu (许峰玮) \inst{1,2} 
\and Ke Wang \inst{1}
\and Tie Liu \inst{3}
\and Lei Zhu \inst{4}
\and Guido Garay \inst{5}
\and Xunchuan Liu \inst{3}
\and Paul Goldsmith \inst{6}
\and Qizhou Zhang \inst{7}
\and Patricio Sanhueza \inst{8,9}
\and Shengli Qin \inst{10}
\and Jinhua He \inst{11,4,5}
\and Mika Juvela \inst{12}
\and Anandmayee Tej \inst{13}
\and Hongli Liu \inst{10}
\and Shanghuo Li \inst{14}
\and Kaho Morii \inst{15,8}
\and Siju Zhang \inst{1}
\and Jianwen Zhou \inst{16}
\and Amelia Stutz \inst{17}
\and Neal J. Evans \inst{18}
\and Kim Kee-Tae \inst{19}
\and Shengyuan Liu \inst{20}
\and Diego Mardones \inst{5}
\and Guangxing Li \inst{21}
\and Leonardo Bronfman \inst{5}
\and Ken'ichi Tatematsu \inst{8,22}
\and Chang Won Lee \inst{23,24}
\and Xing Lu \inst{3}
\and Xiaofeng Mai \inst{3}
\and Sihan Jiao \inst{25}
\and James O. Chibueze \inst{26,27,28}
\and Keyun Su \inst{1}
\and L. Viktor T\'oth \inst{29,30}}

\institute{
Kavli Institute for Astronomy and Astrophysics, Peking University, Beijing 100871, People's Republic of China; {\it fengwei.astro@pku.edu.cn}, {\it kwang.astro@pku.edu.cn} \\
\and
Department of Astronomy, School of Physics, Peking University, Beijing, 100871, People's Republic of China \\
\and
Shanghai Astronomical Observatory, Chinese Academy of Sciences, 80 Nandan Road, Shanghai 200030, People's Republic of China; {\it liutie@shao.ac.cn} \\
\and
Chinese Academy of Sciences South America Center for Astronomy, National Astronomical Observatories, Chinese Academy of Sciences, Beijing, 100101, People's Republic of China; \\
\and
Departamento de Astronom\'ia, Universidad de Chile, Las Condes, 7591245 Santiago, Chile; \\
\and
Jet Propulsion Laboratory, California Institute of Technology, 4800 Oak Grove Drive, Pasadena CA 91109, USA; \\
\and
Center for Astrophysics $|$ Harvard \& Smithsonian, 60 Garden Street, Cambridge, MA 02138, USA; \\
\and
National Astronomical Observatory of Japan, National Institutes of Natural Sciences, 2-21-1 Osawa, Mitaka, Tokyo 181-8588, Japan; \\ 
\and
Astronomical Science Program, The Graduate University for Advanced Studies, SOKENDAI, 2-21-1 Osawa, Mitaka, Tokyo 181-8588, Japan; \\
\and
School of Physics and Astronomy, Yunnan University, Kunming 650091, People's Republic of China; \\
\and
Yunnan Observatories, Chinese Academy of Sciences, 396 Yangfangwang, Guandu District, Kunming, 650216, People's Republic of China; \\
\and 
Department of Physics, University of Helsinki, PO Box 64, FI-00014 Helsinki, Finland; \\
\and 
Indian Institute of Space Science and Technology, Thiruvananthapuram 695 547, Kerala, India; \\
\and
Max Planck Institute for Astronomy, Königstuhl 17, D-69117 Heidelberg, Germany; \\
\and
Department of Astronomy, Graduate School of Science, The University of Tokyo, 7-3-1 Hongo, Bunkyo-ku, Tokyo 113-0033, Japan; \\
\and 
Max-Planck-Institut für Radioastronomie, Auf dem Hügel 69, 53121 Bonn, Germany; \\
\and
Departamento de Astronom´ıa, Universidad de Concepcion, Casilla 160-C, Concepci ´ on, Chile; \\
\and
Department of Astronomy, The University of Texas at Austin, Texas 78712-1205, USA; \\
\and
Korea Astronomy and Space Science Institute, 776 Daedeokdae-ro, Yuseong-gu, Daejeon 34055, Republic of Korea; \\
\and
Institute of Astronomy and Astrophysics, Academia Sinica, Roosevelt Road, Taipei 10617, Taiwan (R.O.C); \\
\and
South-Western Institute for Astronomy Research, Yunnan University, Kunming, People's Republic of China; \\
\and
Astronomical Science Program, The Graduate University for Advanced Studies, SOKENDAI, 2-21-1 Osawa, Mitaka, Tokyo 181-8588, Japan; \\
\and
Korea Astronomy and Space Science Institute, 776 Daedeok-daero, Yuseong-gu, Daejeon 34055, Republic of Korea; \\
\and
University of Science and Technology, 217 Gajeong-ro, Yuseong-gu, Daejeon 34113, Republic of Korea; \\
\and
National Astronomical Observatories, Chinese Academy of Sciences, Beijing 100101, People's Republic of China; \\
\and 
Department of Mathematical Sciences, University of South Africa, Cnr Christian de Wet Rd and Pioneer Avenue, Florida Park, 1709, Roodepoort, South Africa; \\
\and
Centre for Space Research, Physics Department, North-West University, Potchefstroom 2520, South Africa; \\
\and
Department of Physics and Astronomy, Faculty of Physical Sciences, University of Nigeria, Carver Building, 1 University Road, Nsukka 410001, Nigeria; \\
\and
Institute of Physics and Astronomy, E\"otv\"os Lor\`and University, P\'azm\'any P\'eter s\'et\'any 1/A, H-1117 Budapest, Hungary; \\
\and
University of Debrecen, Faculty of Science and Technology, Egyetem t\'er 1, H-4032 Debrecen, Hungary; \\
\vs\no
   {\small Received 2023 month day; accepted 2023 month day}}

\abstract{
Leveraging the high resolution, sensitivity, and wide frequency coverage of the Atacama Large Millimeter/submillimeter Array (ALMA), the QUARKS survey, standing for `Querying Underlying mechanisms of massive star formation with ALMA-Resolved gas Kinematics and Structures', is observing 139 massive star-forming clumps at ALMA Band 6 ($\lambda\sim$ 1.3\,mm). This paper introduces the Atacama Compact Array (ACA) 7-m data of the QUARKS survey, describing the ACA observations and data reduction. Combining multi-wavelength data, we provide the first edition of QUARKS atlas, offering insights into the multiscale and multiphase interstellar medium (ISM) in high-mass star formation. The ACA 1.3\,mm catalog includes 207 continuum sources that are called ACA sources. Their gas kinetic temperatures are estimated using three formaldehyde transitions with a non-LTE radiation transfer model, and the mass and density are derived from a dust emission model. The ACA sources are massive (16--84 percentile values of 6--160\,\msun), gravity-dominated ($M\propto R^{1.1}$) fragments within massive clumps, with supersonic turbulence ($\mathcal{M}>1$) and embedded star-forming protoclusters. We find a linear correlation between the masses of the fragments and the massive clumps, with a ratio of 6\% between the two. When considering the fragments as representative of dense gas, the ratio indicates a dense gas fraction (DGF) of 6\%, although with a wide scatter ranging from 1\% to 10\%. If we consider the QUARKS massive clumps to be what is observed at various scales, then the size-independent DGF indicates a self-similar fragmentation or collapsing mode in protocluster formation. With the ACA data over four orders of magnitude of luminosity-to-mass ratio ($L/M$), we find that the DGF increases significantly with $L/M$, which indicates clump evolutionary stage. We observed a limited fragmentation at the subclump scale, which can be explained by dynamic global collapse process. 
}

   \authorrunning{Xu Fengwei et al.} 
   \titlerunning{ALMA-QUARKS: ACA Continuum Source Catalog}  
   \maketitle

%
%
\section{Introduction} \label{sec:intro}

High-mass stars ($M>8$\,\msun), as the principal sources of UV radiation and heavy elements, play a major role in the evolution of galaxies. However, the properties and evolution of massive clumps hosting the precursors of massive stars currently forming in our Galaxy are still poorly known \citep[e.g.,][]{Zinnecker2007MSF,Tan2014MSF,Pineda2023PP7}. Fragmentation of massive clumps into dense cores, where star formation ultimately occurs, is a critical step in the mass assembly process that gives rise to stars and clusters, as highlighted by \citet{Motte2018Review}. Investigating how dense gas is concentrated and structured within these massive clumps serves as an intermediate step in understanding this intricate process. 

Recent studies, such as \citet{Peretto2020ClumpFed}, have conducted sub-millimeter continuum surveys of infrared dark clouds, revealing that the evolution of massive compact sources in mass-versus-temperature diagrams is better explained by an accretion scenario where cores gain mass while simultaneously collapsing to form protostars. Furthermore, the findings from \citet{Rigby2021GASTON} provide evidence for the mass growth of clumps, suggesting that similar mass accumulation processes may occur on a broader range of physical scales, which is further verified in several multiscale case studies \citep[e.g.,][]{Neupane2020Global,Xu2023ATOMS-XV,Yang2023G310}.
On the simulation side, the mass growth of a `core' is believed to be the result of the collapse of the surrounding parsec-scale mass reservoir called a `clump', hence the accretion scenario described above is referred to as `clump-fed' \citep{Wang2010ClumpFed}. 

The `clump-fed' scenario suggests that there must be a link between the properties of a clump and the fragments that form within it. In the case of clump fragmentation, \citet{Lin2019SABOCA} find a correlation between the mass of the clump and the mass of its most massive fragment of massive clumps from ``the APEX Telescope Large Area Survey of the Galaxy'' \citep[ATLASGAL;][]{Schuller2009ATLASGAL}. Besides, \citet{Barnes2021IRDCs} also find that more massive, and more turbulent clouds make more $\sim0.1$\,pc scale cores. \citet{Anderson2021HFS} collected a sample of massive clumps with a wide range of evolutionary stages, and suggested time-dependent correlation between clump and core mass especially in hub-filament systems. The relation between the formation of dense cores and the properties of clumps such as turbulence is also discussed \citep[e.g.,][]{Xu2021HGaL,Xu2024Scarcity}. On a smaller scale, studies such as \citet{Palau2014Fragmentation,Palau2021Fragmentation} find correlations between the level of fragmentation within massive dense cores ($<0.1$\,pc) and their average volume density, aligning with the expectations of the Jeans instability \citep{Sanhueza2019ASHES,Morii2024}. Furthermore, using ALMA with a resolution of approximately 0.02\,pc, \citet{Xu2024ASSEMBLE} identified a sublinear correlation between the mass of the clump and the mass of its most massive core, in a sample of 11 massive protoclusters that show clump-scale infall motion. For comparison, no such correlation was found in a sample of 39 massive early-stage clumps from \citet{Morii2023ASHES}. This suggests that the mass correlation between the clumps and the cores gradually builds up over the evolution of massive clumps. 

Despite previous great advances, our understanding of the formation process of massive stars remains unclear and divided due to observational difficulties. On the one hand, considering their large distances (a few kpc) and high dust extinction, studies of massive clumps need high-resolution interferometric observations to resolve their internal gas structures and kinematics \citep{Wang2015Book,Motte2018Review,LinY2021PhDThesis}. On the other hand, obtaining robust and definitive conclusions regarding the accretion history of high-mass stars requires a larger statistical sample. This, in turn, calls for rapid survey capabilities with adequate sensitivity. The Atacama Large Millimeter/submillimeter Array (ALMA), with both high resolution and high sensitivity, offers a unique valuable opportunity to investigate the hierarchical structures in massive star-forming regions in great detail. Therefore, we performed a 1.3-mm ALMA survey called ``Querying Underlying mechanisms of massive star formation with ALMA-Resolved gas Kinematics and Structures'' (QUARKS; PIs: Lei Zhu, Guido Garay and Tie Liu; Project ID: 2021.1.00095.S). The details of the survey description and the showcase of data combinations can be found in \citet{Liu2023QUARKS}. 

In this paper, we focus mainly on the Atacama Compact Array (ACA) 7-m continuum and line data sets. With relatively little free-free emission contamination and a maximum recoverable scale as large as $\sim27$\arcsec, ACA 1.3\,mm continuum data are useful for tracking dense molecular gas within massive clumps. With an angular resolution of $\sim5$\arcsec, equivalent to 0.07\,pc at a typical distance of 3\,kpc within the QUARKS sample, the ACA data provide a global view of massive protostars or protoclusters therein. We first introduce the ACA observations and data imaging in Section\,\ref{sec:data}, and then provide the first edition of the QUARKS atlas in Section\,\ref{sec:atlas}. Section\,\ref{sec:results} constructs the ACA 1.3\,mm continuum source catalog including physical parameters. In Section\,\ref{discuss:nature}, we discuss the physical nature of ACA sources and find that they are condensed gas fragments within massive clumps. In Section\,\ref{discuss:corr}, we find a mass correlation between clumps and their fragments. In Section\,\ref{discuss:DGF}, we discuss the size-invariant and time-variant dense gas fraction. In Section\,\ref{discuss:fragmentation}, we discuss limited fragmentation at the subclump scale. In Section\,\ref{sec:sum}, we present a brief summary. 

\section{QUARKS ACA Data} \label{sec:data}

\subsection{Observing Strategy}

QUARKS acts as a follow-up 1.3-mm survey of ALMA Three-millimeter Observations of Massive Star-forming regions \citep[ATOMS;][]{Liu2020ATOMS-I}, and aims at studying even smaller structures within the 3-mm cores or core clusters within massive star-forming clumps. To ensure uniformity of the sample and solid detection at 1.3\,mm, we exclude: 1) two low-mass clumps ($<15$ $M_\odot$); 2) four sources dominated by extended \hii~regions with angular sizes larger than the primary beam at Band 6; 3) one source without continuum emission detection by ATOMS. As a result, 139 ATOMS massive clumps are selected as the QUARKS sample, hereafter the QUARKS clumps. A total of 156 ALMA 1.3\,mm pointings were performed because some 3\,mm emission show elongated or extended morphologies and need two mosaicked pointings (dual-pointing mosaicked field). The on-source time of each pointing was about 5 minutes. 

The QUARKS ACA observations are separated into 15 scheduling blocks (SBs), hereafter called groups for short. The group ID and the number of fields therein are listed in columns (1)--(2) of Table\,\ref{tab:almaobs}. To finish the observing queue, 1--3 execution blocks (EBs) were performed on different observing dates, which are listed in column (3). Fields in the same group have the proximity of sky coordinates and the same EBs, so they share similar minimum and maximum baselines (BL), angular resolution (AR), and maximum recoverable scale (MRS), which are listed in columns (4)--(6). Variations in AR and MRS are mostly the result of different configuration of the array and source elevation. Phase and flux calibrators are listed in column (7), while bandpass and flux calibrators are listed in column (8). 

The observations utilized ALMA Band 6 receivers in dual-polarization mode, with the correlator frequencies configured into four spectral windows (SPWs 1--4). The four SPWs were designed with center frequencies at approximately 217.92\,GHz, 220.32\,GHz, 231.37\,GHz, and 233.52\,GHz, each with a bandwidth of 2\,GHz and 4096 channels. This setup was chosen to cover a wide range of commonly used tracers representing different environments and excitation conditions. The targeted lines included, but were not limited to: 1) outflow tracers (e.g., CO, $^{13}$CO, SiO, SO, H$_2$CO); 2) cold gas tracer (N$_2$D$^+$); 3) dense core and filament tracers (e.g., C$^{18}$O, HC$_3$N); 4) hot core tracers (e.g., CH$_3$OH, C$_2$H$_5$CN, NH$_2$CHO, CH$_3$CN); 5) ionized-gas/\hii-region tracer (H$_{30}\alpha$). The basics of the main targeted lines are summarized in Table 2 of \citet{Liu2023QUARKS}. 

\begin{table*}[!thb]
\centering
\caption{QUARKS ACA Observation and Imaging Result Logs \label{tab:almaobs}}
\setcellgapes{2pt}
\renewcommand{\arraystretch}{1.8} 
\begin{tabular}{ccccccccc}
\hline
\hline
Group ID & $n_{\rm field}$$^{a}$ & Obs. Date & Min./Max. BL & AR & MRS & \multicolumn{2}{c}{Calibrators} & cont. rms \\
\cline{7-8}
& & & (m/m) & (\arcsec) & (\arcsec) & Phase & Bandpass/Flux & (\mjybeam) \\
(1) & (2) & (3) & (4) & (5) & (6) & (7) & (8) & (9) \\
\hline
1 & 6 & 2021-11-06 & 8.9/45.0 & 4.9 & 20.0 & J0922-3959 & J1058+0133 & 10 \\
\hline
\multirow{2}{*}{\makecell[c]{2}} & \multirow{2}{*}{\makecell[c]{3}} & 2021-10-22 & 8.9/44.7 & 5.1 & 28.3 & J1047-6217 & J1058+0133 & \multirow{2}{*}{\makecell[c]{6}} \\
 &  & 2022-05-21 & 8.9/45.0 & 4.9 & 28.7 & J1047-6217 & J1058+0133 & \\
\hline
3 & 7 & 2023-01-02 & 8.9/48.9 & 5.0 & 28.3 & J1424-6807 & J1427-4206 & 9 \\
\hline
\multirow{2}{*}{\makecell[c]{4}} & \multirow{2}{*}{\makecell[c]{12}} & 2023-01-13 & 8.9/48.9 & 4.8 & 28.3 & J1337-6509 & J1427-4206 & \multirow{2}{*}{\makecell[c]{7}} \\
 &  & 2023-01-16 & 8.9/48.0 & 5.0 & 28.3 & J1337-6509 & J1427-4206 & \\
\hline
\multirow{3}{*}{\makecell[c]{5}} & \multirow{3}{*}{\makecell[c]{24}} & 2023-04-09 & 8.9/48.0 & 4.5 & 21.1 & J1604-4441 & J1427-4206 & \multirow{3}{*}{\makecell[c]{10}} \\
 &  & 2023-04-13 & 8.9/48.0 & 4.9 & 27.1 & J1617-5848 & J1427-4206 & \\
 &  & 2023-04-18 & 8.9/48.0 & 4.9 & 27.1 & J1617-5848 & J1427-4206 & \\
\hline
6 & 5 & 2023-01-01 & 8.9/48.9 & 5.3 & 28.3 & J1524-5903 & J1427-4206 & 18 \\
\hline
\multirow{2}{*}{\makecell[c]{7}} & \multirow{2}{*}{\makecell[c]{13}} & 2023-05-17 & 8.9/48.9 & 4.8 & 27.1 & J1733-3722 & J1924-2914 & \multirow{2}{*}{\makecell[c]{9}} \\
 & & 2023-05-20 & 8.9/48.9 & 4.8 & 27.1 & J1733-3722 & J1924-2914 & \\
 \hline
8 & 3 & 2023-01-14 & 8.9/48.0 & 4.9 & 21.1 & J1924+1540 & J2232+1143 & 43 \\
\hline
9 & 3 & 2023-01-22 & 8.9/48.0 & 5.0 & 28.3 & J1744-3116 & J1427-4206 & 22 \\
\hline
\multirow{2}{*}{\makecell[c]{10}} & \multirow{2}{*}{\makecell[c]{8}} & 2023-03-04 & 9.1/45.0 & 4.9 & 21.1 & J1851+0035 & J1924-2914 & \multirow{2}{*}{\makecell[c]{5}} \\
 & & 2023-04-16 & 8.9/48.0 & 4.9 & 27.1 & J1851+0035 & J1924-2914 & \\
\hline
11 & 7 & 2023-04-08 & 8.9/48.0 & 4.5 & 21.1 & J1717-3342 & J1427-4206 & 28 \\
\hline
\multirow{2}{*}{\makecell[c]{12}} & \multirow{2}{*}{\makecell[c]{18}} & 2023-04-25 & 8.9/48.0 & 4.9 & 27.1 & J1604-4441 & J1427-4206 & \multirow{2}{*}{\makecell[c]{5}} \\
 & & 2023-05-01 & 8.9/48.0 & 4.9 & 27.1 & J1604-4441 & J1427-4206 & \\
\hline
\multirow{2}{*}{\makecell[c]{13}} & \multirow{2}{*}{\makecell[c]{11}} & 2023-03-04 & 9.1/45.0 & 4.9 & 21.1 & J1851+0035 & J1924-2914 & \multirow{2}{*}{\makecell[c]{8}} \\
 & & 2023-04-09 & 8.9/48.0 & 4.5 & 21.1 & J1851+0035 & J1924-2914 & \\
 \hline
14 & 6 & 2023-04-21 & 8.9/48.0 & 4.9 & 27.1 & J1820-2528 & J1924-2914 & 17 \\
\hline
\multirow{2}{*}{\makecell[c]{15}} & \multirow{2}{*}{\makecell[c]{13}} & 2023-04-17 & 8.9/48.0 & 4.9 & 27.1 & J1832-2039 & J1924-2914 & \multirow{2}{*}{\makecell[c]{12}} \\
 & & 2023-04-18 & 8.9/48.0 & 4.9 & 27.1 & J1832-2039 & J1924-2914 & \\
\hline
\hline
\end{tabular}{
\begin{flushleft}
The QUARKS ACA observations are separated into 15 scheduling blocks which are called groups for short. The group ID, the number of targets therein, and the dates of observation are listed in columns (1)--(3). The minimum and maximum baselines (BL) of the configuration array are listed in column (4). Group-averaged angular resolution (AR) and maximum recoverable scale (MRS) are listed in columns (5)--(6). The phase calibrator(s) are listed in column (7) while the bandpass and flux calibrators are listed in column (8). The aggregated continuum imaging rms and spectral line rms per channel are listed in column (9). \\
$^{(a)}$Including both single-pointing and dual-pointing fields, therefore 139 in total.
\end{flushleft}
}
\end{table*}

\subsection{Data Calibration and Imaging}

QUARKS ACA data were acquired during the ALMA Cycle 8 and 9 observations. The data were routinely calibrated using the ALMA pipelines \footnote{\href{https://almascience.nrao.edu/processing/science-pipeline}{https://almascience.nrao.edu/processing/science-pipeline}} of Common Astronomy Software Applications \citep[CASA;][]{McMullin2007CASA,CASA2022} in the corresponding versions of 6.2.1 and 6.4.1. The frequency tunings of the correlator are four wide bands, each with 2\,GHz separated by 4096 channels. The edge channels of each spectral window ($\sim2\times128$) were flagged in the first version of data release due to the high temperature of the system noise and therefore the high level of noise. 

Line emission channels were flagged to obtain the continuum and spectral lines simultaneously. \citet{Liu2023QUARKS} identified all the transitions of strong lines within the four SPWs by matching the reduced data cubes of the ALMA pipeline and the laboratory databases for the spectral lines \citep[CDMS;][]{Muller2001CDMS}. The QUARKS team generated a model spectrum as a mask for line emission channels. For each source, the model spectrum was shifted and expanded with a width of 50\,\kms~to ensure clean channels free from multiple velocity components and spectral line wings. With this method the line emission channels were flagged and the line-free channels were subtracted from the four spectral windows in the Fourier space using the task \textit{uvcontsub} with linear fitting (polynomial order of 1). The continuum and line cube imaging processes were performed by the task \textit{tclean} in CASA 6.5.6, with \texttt{briggs} robust weighting of 0.5. In the cleaning process, the images/cubes were automatically masked by \texttt{auto-multithresh} algorithm whose input parameters are recommended by the official guides \footnote{\href{https://casaguides.nrao.edu/index.php?title=Automasking_Guide}{casaguide:automask}} for the ACA data. At the beginning of each minor cycle, the cleaning mask was updated on the basis of the current residual image. The algorithm uses multiple thresholds based on the noise and sidelobe levels in the residual image to determine the cleaning mask. Within the cleaning mask, we set the stopping \texttt{noisethreshold} to be 5$\sigma$ and \texttt{sidelobethreshold} to be $1.25\sigma$ in the dirty image/cube. However, the cleaning algorithm diverges when some fields have relatively strong emission or side lobes at the edge. For these fields, we performed a manual mask to further improve the imaging fidelity. To recover the potential large-scale structures in the spectral lines and mitigate artifacts produced by extended emission, we applied the \texttt{multiscale} deconvolver \citep{Cornwell2008Multiscale} to the cleaning process of line cubes, with scales of [0,5,15]. We uniformly set the image size of $108\times108$\,pix$^2$ and cell size of 1\arcsec, to fully cover the 17 mosaic fields with dual pointings. The primary beam correction was performed with \texttt{ pblimit = 0.2}. 

The noise rms of the cleaned continuum image is tabulated in column (9) of Table\,\ref{tab:almaobs}. The rms levels have large variations between groups because some sources are too strong and the continuum sensitivities are limited by the dynamic range. Self-calibration can improve the sensitivity, but will not contribute much to the science goals in this paper. Therefore, self-calibration has not been performed on our released ACA data. 

\section{QUARKS Atlas} \label{sec:atlas}

With the inclusion of QUARKS as a high-resolution submillimeter dataset, we are now equipped to construct a comprehensive data atlas for the QUARKS sample, offering insights into the multiscale and multiphase interstellar medium (ISM) in high-mass star formation. 

\begin{figure*}[!ht]
\centering
\includegraphics[width=\linewidth]{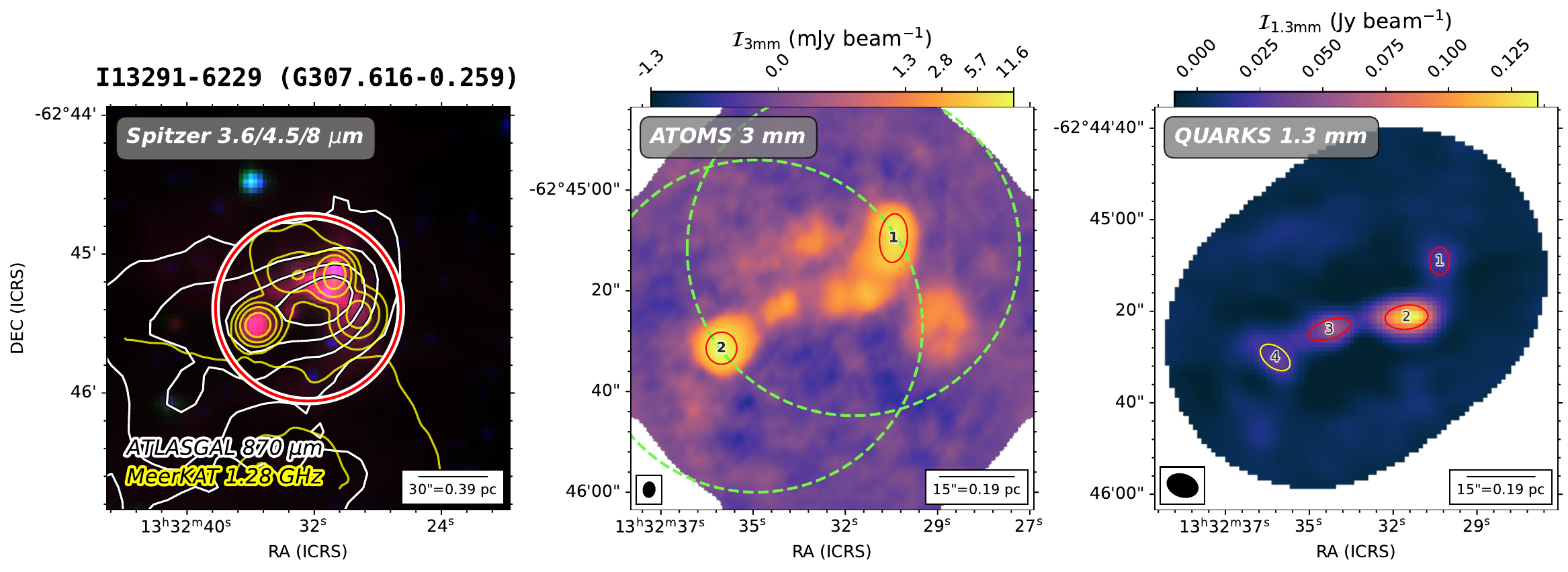}
\caption{QUARKS multi-band atlas of representative source I13291-6229. \textit{Left panel}: the background is the Spitzer 3.6/4.5/8\,$\mu$m pseudo color map, overlaid with Herschel 500\,$\mu$m (white contours) and MeerKAT Galactic Plane Survey (MGPS) 1.28\,GHz data (yellow contours). The red circle indicates the field of view ($\sim80$\arcsec) of the combined ATOMS 12m + ACA 3\,mm continuum data. \textit{Middle panel}: the background is the ATOMS combined 3\,mm continuum data, linearly scaled from $-9\sigma$ to $9\sigma$ and logarithmically scaled from $9\sigma$ to peak intensity. The source IDs are in order from North to South, and the nomenclature follows ``\#Field\_ATOMS\#ID''. The green dashed circle(s) indicate the QUARKS pointing(s), with size of 7-m primary beam response of 0.2. The ATOMS beam size is shown on the bottom left. \textit{Right panel}: the background is the QUARKS ACA 1.3\,mm continuum data, linearly scaled from $-3\sigma$ to peak intensity. The continuum sources are shown as red ellipses (SNR$>9$) and yellow ellipses (SNR$<9$). The source IDs are in order from North to South and the nomenclature follows ``\#Field\_ACA\#ID''. The QUARKS beam size is shown in the bottom left. The scale bars in three panels are shown on the bottom right.}
\label{fig:atlas_example}
\end{figure*}

The mid infrared (MIR) Spitzer facility was equipped with the InfraRed Array Camera (IRAC) instrument that provided images at 3.6, 4.5, 5.8, and 8.0\,$\mu$m simultaneously. Here we combine the Galactic Legacy Infrared Mid-Plane Survey Extraordinaire (GLIMPSE) data at 3.6, 4.5, and 8.0\,$\mu$m into the pseudo color map in the left panel of Figure\,\ref{fig:atlas_example}. The three bands can provide information about young stellar objects (YSOs), shocked molecular gas in protostellar outflows, and hot dust or polycyclic aromatic hydrocarbon (PAH) emission, respectively. The angular resolution of the images is smoothed to a uniform value of $2.0$\arcsec. Some of our source positions have no available Spitzer data, and for these we use ALLWISE data at 3.4, 4.6, and 22.0\,$\mu$m \citep{Wright2010WISE,Mainzer2011NOEWISE} instead, which have an angular resolution of 6\parcsec1, 6\parcsec4, 6\parcsec5, and 12\parcsec0, respectively.

The submillimeter emission traces the cold and dense gas well. ATLASGAL explores the inner Galactic plane at submillimeter wavelengths ($\sim870$\,$\mu$m) with a beam size of 19\parcsec2 \citep{Schuller2009ATLASGAL}. We use the ATLASGAL data to trace the large scale cold and dense gas which indicates the star formation region, in the white contours on the left panel. If no ATLASGAL data are available, far infrared (FIR) Herschel data from ``the Herschel Infrared Galactic Plane Survey'' \citep[HiGAL;][]{Traficante2011HiGAL} at 500\,$\mu$m is used instead, with a nominal beam size of 34\parcsec5 \citep{Traficante2011HiGAL}. 

The MeerKAT Galactic Plane Survey 1.28\,GHz data \citep[MGPS;][]{Padmanabh2023MeerKAT,Goedhart2023MGPS} provides essential radio information with a resolution of $\sim8\arcsec$. It is overlaid with yellow contours to indicate the ionized gas from ultra-compact \hii~(UC\hii) and hyper-compact \hii~(HC\hii) regions as well as radio jets. 

On a smaller scale, the ATOMS 12m+ACA combined 3\,mm continuum data can trace both the dust emission from cold dense cores and/or the free-free emission from UC\hii~and HC\hii~regions \citep{Liu2021ATOMS-III}. The potential contamination from free-free emission can be estimated using centimeter wavelength data \citep[and forthcoming ATCA project]{Avison2015SDC335,Olguin2022SDC335,Xu2023ATOMS-XV}. 

In Figure\,\ref{fig:atlas_example}, the red circle indicates the field of view ($\sim80$\arcsec) of the ATOMS continuum data, zooming in on the substructures of $\sim2\arcsec$ in one of the QUARKS massive clump I13291-6229, as shown in the middle panel. The 3\,mm continuum emission of I13291-6229 exhibits a filamentary morphology from the southeast to the northwest. Dense cores identified in \citet{Liu2021ATOMS-III} are marked with red ellipses and ID numbers. As a follow-up, the QUARKS 1.3\,mm ACA observations are pointed towards the 3\,mm continuum emission region, as marked by the white dashed circles where the 7-m primary beam response is 0.2. In the right panel, the QUARKS ACA data are shown in the background color map and four 1.3\,mm sources are identified, where three are identified as solid detection with SNR$>9$ in red and one with SNR$<9$ in yellow. 

The QUARKS survey updates the clump distances using the H$^{13}$CO$^+$ lines of the ATOMS survey and the latest model for the rotation curve of the Milky Way \citep{Reid2019Distance}, as listed in Table A. of \citet{Liu2023QUARKS}. At the bottom right of Figure\,\ref{fig:atlas_example}, both the angular and physical scale bars are shown, with updated clump distance. The complete QUARKS data atlases are presented in the supplementary material. 

\section{Results} \label{sec:results}

\subsection{ACA 1.3 mm Continuum Source Catalog} \label{results:sources}

We adopted an automatic source extraction algorithm \textsc{SExtractor} \footnote{\href{https://sextractor.readthedocs.io/en/latest/Introduction.html}{https://sextractor.readthedocs.io/en/latest/Introduction.html}.} \citep{Bertin1996SExtractor} on the ACA 1.3\,mm continuum emission maps to extract the 1.3\,mm continuum sources. The advantages of \textsc{SExtractor} in our case are: 1) to subtract the background diffuse emission and rms noise automatically; 2) to support local rms noise input as pixelwise thresholds; 3) to deblend the potentially overlapped sources in crowded fields. The details for the algorithm input parameters are described in Appendix\,\ref{app:sextractor}. 

As a result, a total of 207 ACA 1.3\,mm continuum sources are extracted from the 139 massive star-forming clumps and the fundamental measurements are summarized in Table\,\ref{tab:measurements}. At least one and at most five sources have been detected in one clump. The field name is listed in column (1) and the continuum source ID which is given in order from North to South is listed in column (2). Hereafter, the format of ACA 1.3\,mm continuum source obeys ``\#Field\_ACA\#ID''. The ICRS coordinates of the sources are listed in columns (3)--(4). The FWHM of the major and minor axes and the position angle are listed in columns (5)--(7). The integrated flux and peak intensity, which are corrected by primary beam (see details in Section\,\ref{app:sextractor}), are listed in columns (8)--(9). The signal-to-noise (SNR) ratio, defined as the peak intensity over the local rms, is listed in column (10). 

We note that the maximum observed angular size (convolved with the beam) of ACA sources is $\sim14\arcsec$, which is only half of the maximum recoverable scale in most cases (see Table\,\ref{tab:almaobs}). We thus ignore the effects of missing flux in the following analyses. 

\begin{table*}[!thb]
\centering
\caption{Basic Measurements of ACA 1.3\,mm Continuum Sources \label{tab:measurements}.}
\renewcommand{\arraystretch}{1.5} 
\begin{tabular}{cccccccccc}
\hline
\hline
Field & ID & \multicolumn{2}{c}{Equatorial Coordinates} & \multicolumn{2}{c}{$\theta_{\rm maj}\times\theta_{\rm min}$} & PA & $F_{\rm int}$ & $I_{\rm peak}$ & SNR \\
& ACA & RA (ICRS) & DEC (ICRS) & (arcsec) & (arcsec) & (deg) & (Jy) & (\jybeam) & \\
\cmidrule(r){3-4} \cmidrule(r){5-6}
(1) & (2) & (3) & (4) & (5) & (6) & (7) & (8) & (9) & (10) \\
\hline
I08303-4303 & 1 & 08:32:08.7 & -43:13:46.2 & 10.1 & 7.0 & 128.6 & 0.811 & 0.347 & 31.8 \\
I08448-4343 & 1 & 08:46:35.1 & -43:54:22.6 & 6.8 & 5.1 & 64.9 & 0.143 & 0.108 & 11.4 \\
I08448-4343 & 2 & 08:46:34.7 & -43:54:33.5 & 7.6 & 4.1 & 156.6 & 0.087 & 0.076 & 9.5 \\
I08448-4343 & 3 & 08:46:33.4 & -43:54:36.5 & 7.5 & 4.6 & 5.9 & 0.171 & 0.126 & 16.5 \\
I08448-4343 & 4 & 08:46:32.5 & -43:54:37.1 & 12.2 & 4.4 & 5.2 & 0.175 & 0.114 & 16.5 \\
I08470-4243 & 1 & 08:48:47.8 & -42:54:26.4 & 10.0 & 6.0 & 74.6 & 0.989 & 0.623 & 77.2 \\
I09002-4732 & 1 & 09:01:54.3 & -47:44:09.8 & 7.2 & 5.8 & 2.6 & 2.675 & 1.720 & 31.2 \\
I09018-4816 & 1 & 09:03:33.3 & -48:28:01.0 & 12.1 & 7.0 & 96.9 & 1.403 & 0.617 & 40.5 \\
I09094-4803 & 1 & 09:11:08.6 & -48:15:44.1 & 5.3 & 5.0 & 164.7 & 0.063 & 0.056 & 14.3 \\
I09094-4803 & 2 & 09:11:08.3 & -48:15:53.3 & 6.6 & 5.2 & 81.4 & 0.093 & 0.073 & 22.3 \\
\hline
\end{tabular}
\begin{flushleft}
The field name and the continuum source ID are listed in columns (1)--(2). The ICRS coordinates of the barycenter are listed in columns (3)--(4). The FWHM of the major and minor axes ($\theta_{\rm maj}$ and $\theta_{\rm min}$), and the position angle (PA) of sources are listed in columns (5)--(7). The integrated flux $F_{\rm int}$ and the peak intensity $I_{\rm peak}$ are listed in columns (8)--(9). The signal-to-noise ratio (SNR) is listed in column (10). The table is available in its entirety in machine-readable form. 
\end{flushleft}
\end{table*}

\subsection{Temperature Estimation from Formaldehyde} \label{results:temperature}

Formaldehyde (\htco) is a suitable spectroscopic tool to derive the kinetic temperature of the molecular gas \citep{Ao2013CMZTemp}. For instance, observations throughout the Central Molecular Zone \citep[CMZ;][]{Ginsburg2016CMZTemp}, show that the \htco~emission is spatially widespread and correlated with dust emission. Formaldehyde has two isomeric species, ortho-\htco~(o-\htco) and para-\htco~(p-\htco). \citet{Kahane1984OPR} reported that p-\htco~is 1--3 times less abundant than o-\htco~in three low-mass star forming regions. \citet{Mangum1993H2COTemp} reported ortho-to-para ratio of 1.5--3 in the Orion-KL high-mass star-forming region. Therefore, temperature estimation from p-\htco~transition lines minimizes the uncertainties from a high optical depth. For example, \citet{Tang2017H2CO} find that the kinetic temperatures derived from p-\htco~show a good agreement with those of the dust in the Large Magellanic Cloud (LMC), suggesting that the dust and p-\htco~molecules are probing the same (or similar) gas component. More importantly, \citet{Tang2021H2CO} mapped two massive star forming regions in the Large Magellanic Cloud (LMC) using ALMA at a resolution of 0.4\,pc and find a consistency in the spatial distribution of the dense gas traced by p-\htco~with that of the 1.3\,mm dust. Above all, these studies provide a practical foundation for using the p-\htco~molecule to estimate the dust temperature of the ACA sources. 

The QUARKS frequency tunings in SPW1 are designed to cover the p-\htco~transition triplet, i.e., $J_{\rm K_A,K_c} = 3_{0,3} \rightarrow 2_{0,2}$ ($3_{03}$--$2_{02}$) at 218.22219\,GHz, $3_{2,2}\rightarrow2_{2,1}$ ($3_{22}$--$2_{21}$) at 218.47563\,GHz, and $3_{2,1}\rightarrow2_{2,0}$ ($3_{21}$--$2_{20}$) at 218.76007\,GHz. Their upper state energies are 21.0\,K, 68.09\,K, and 68.11\,K, respectively. We extracted spectra from 218.1 to 281.9\,GHz in SPW1 to fully cover the frequency range of the \htco~triplet, and perform a five-parameter model using the formaldehyde official model\footnote{\href{https://pyspeckit.readthedocs.io/en/latest/formaldehyde_model.html}{https://pyspeckit.readthedocs.io/en/latest/formaldehyde\_model.html}} by \textsc{pyspeckit} \citep{Ginsburg2011Pyspeckit,Ginsburg2022Pyspeckit} with a non-LTE RADEX model. The details of the fitting model and results are summarized in Appendix\,\ref{app:h2co_fit}. The derived kinetic temperatures have values between 24 to 180\,K with median of 72\,K, which are listed in column (3) of Table\,\ref{tab:physical}. 

\begin{figure*}
\centering
\includegraphics[width=0.8\linewidth]{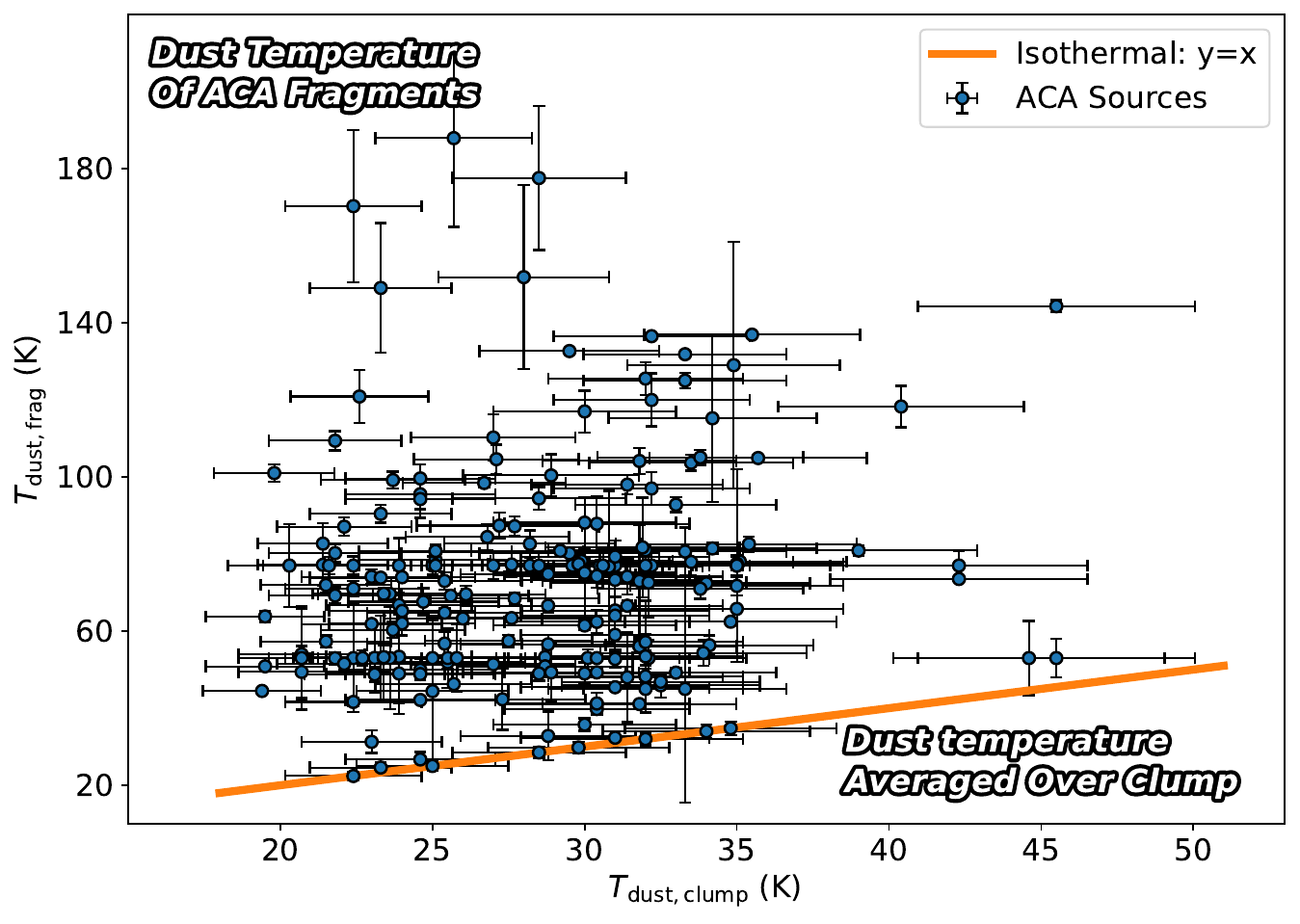}
\caption{Dust temperature at the scales of clump ($T_{\rm dust,clump}$) and embedded ACA fragment ($T_{\rm dust,frag}$). $T_{\rm dust,frag}$ is derived from the non-LTE \htco~modeling based on the assumption that dust temperature is identical to gas temperature. $T_{\rm dust,clump}$ is derived from infrared dust emission SED fitting. The ``isothermal line'' with $T_{\rm dust,frag}=T_{\rm dust,clump}$ is shown in orange. The uncertainty of $T_{\rm dust,frag}$ is given by model fitting and the relative uncertainty of $T_{\rm dust,clump}$ is set to be 10\% for all sources. \label{fig:temp}}
\end{figure*}

A mixture of gas and dust having density over $10^{4.5}$\,cm$^{-3}$ due to collisional process \citep{Goldsmith2001Coolant} yields $T_{\rm dust}$ equal to $T_{\rm kin}$, which holds for most of the ACA sources (refer to Section\,\ref{nature:dense}). Therefore, in this work, we assume that the dust temperature of ACA source/fragment ($T_{\rm dust,frag}$) equals to the kinetic temperature derived from H$_2$CO triplet line fitting ($T_{\rm kin,H_2CO}$). 

Figure\,\ref{fig:temp} illustrates the relationship between fragment-scale dust temperature $T_{\rm dust,frag}$ and the clump-averaged dust temperature $T_{\rm dust,clump}$ determined through infrared-to-submillimeter SED fitting \citep[see method in][]{Urquhart2018ATLASGAL}. All the data points lie above the ``isothermal line'' ($T_{\rm kin,H_2CO}=T_{\rm dust,clump}$), shown in orange color. This suggests an outward negative temperature gradient, with the inner dense gas traced by H$_2$CO being warmer than the surrounding low-density gas and warmer than gas on average. The temperature gradient serves as an indirect observational evidence of the idea that the ACA sources are active star-forming regions. We note that seven sources with failed fitting (six with a strong self-absorption line profile and one with a weak detection) are assigned with the same value as $T_{\rm dust,clump}$, therefore on the orange line in Figure\,\ref{fig:temp}. 

\subsection{Physical Parameters of Sources} \label{results:physics}

Assuming that all the emission comes from dust in a single component with $T_{\rm dust}$ and that the dust emission is optically thin, the masses of the sources can be calculated using
\begin{equation}\label{eq:mass}
   M_{\rm source} = \mathcal{R}\frac{F_{\rm int} D^2}{\kappa_\nu B_\nu (T_{\rm dust})},
\end{equation}
where $F_{\rm int}$ is the measured integrated flux of dust emission, $\mathcal{R}$ is the gas-to-dust mass ratio (assumed to be 100), $D$ is the clump distance, $\kappa_{\nu}$ is the dust opacity per gram of dust, and $B_\nu (T_{\rm dust})$ is the Planck function at a given dust temperature $T_{\rm dust}$. In our case, $\kappa_{\nu}$ is assumed to be 1\,cm$^2$\,g$^{-1}$ at $\nu\sim230$\,GHz which is interpolated from the given table in \citet{O&H1994dust}, assuming grains with thin ice mantles and the size distribution given by \citet{MRN1997} and a typical gas density of $10^6$\,cm$^{-3}$ in our sample. Substituting $T_{\rm dust}$ assumed above to Equation\,\ref{eq:mass}, the ACA source masses are then calculated and listed in the column (4). The major sources of uncertainty in the mass calculation come from the gas-to-dust ratio and the dust opacity. We adopt the uncertainties derived by \citet{Sanhueza2017G28} of 28\% for the gas-to-dust ratio and of 23\% for the dust opacity, contributing to the $\sim36$\% uncertainty of the specific dust opacity. The uncertainty of $F_{\rm int}$ from flux calibration (assumed to be 10\%; Yun et al. 2022 \footnote{\href{https://library.nrao.edu/public/memos/alma/main/memo620.pdf}{ALMA Memo 211}}) and the uncertainty of distance (assumed to be 20\%) are included. Monte Carlo methods are adopted for uncertainty estimation and $1\sigma$ confidence intervals are given. 

ACA sources are characterized by 2D Gaussian-like ellipses with the FWHM of the major and minor axes ($\theta_{\rm maj}$ and $\theta_{\rm min}$), and position angle (PA) listed in columns (5)--(6) of Table\,\ref{tab:measurements}. Following \citet{Rosolowsky2010BGPSII} and \citet{Contreras2013ATLASGAL,Urquhart2014ATLASGAL}, the source angular size can be calculated as the geometric mean of the deconvolved major and minor axes.
\begin{equation}\label{eq:theta}
    \theta_{\rm dec} = \eta\left[\left(\sigma^2_{\rm maj}-\sigma^2_{\rm bm}\right)\left(\sigma^2_{\rm min}-\sigma^2_{\rm bm}\right)\right]^{1/4},
\end{equation}
where $\sigma_{\rm maj}$ and $\sigma_{\rm min}$ are calculated from $\theta_{\rm maj}/\sqrt{8\ln2}$ and $\theta_{\rm min}/\sqrt{8\ln2}$ respectively. 
$\sigma_{\rm bm}$ is the averaged dispersion size of the beam (i.e., $\sqrt{\theta_{\rm bmaj}\theta_{\rm bmin}/(8\ln2)}$ where $\theta_{\rm bmj}$ and $\theta_{\rm bmin}$ are the FWHM of the major and minor axis of the beam). $\eta$ is a factor that relates the size of the emission distribution dispersion to the determined angular radius of the object. $\eta=2.4$, the median value derived for a range of models consisting of a spherical emissivity distribution \citep{Rosolowsky2010BGPSII}, is adopted here. Therefore, the physical size can be calculated directly using $R_{\rm dec} = \theta_{\rm dec}\times D$, as shown in column (5) of Table\,\ref{tab:physical}. 

The source peak column density is estimated from
\begin{equation}
    N^{\rm peak}_{\ssstyle \rm H_2} = \mathcal{R} \frac{I_{\rm peak}}{\Omega\mu_{\ssstyle \rm H_2} m_{\ssstyle \rm H}\kappa_{\nu}B_{\nu}(T_{\rm dust})},
\end{equation}
where $I_{\rm peak}$ is the measured peak flux of source within the beam solid angle $\Omega$. 

The surface density averaged by the source can be calculated by $\Sigma=M_{\rm source}/(\pi R^2_{\rm dec})$. 
The source-averaged number density, $n_{\rm H_2}$, is then calculated by assuming a spherical source,
\begin{equation} \label{eq:volume_density}
    n_{\ssstyle \rm H_2} = \frac{M_{\rm source}}{(4/3) \pi \mu_{\ssstyle \rm H_2} m_{\ssstyle \rm H}R_{\rm dec}^3},
\end{equation}
where $\mu_{\ssstyle \rm H_2}$ is the molecular weight per hydrogen molecule and $m_{\rm H}$ is the mass of a hydrogen atom. Throughout the paper, we adopt the molecular weight per hydrogen molecule $\mu_{\ssstyle \rm H_2} = 2.81$ \citep{Evans2022SlowSF}. The calculated peak column density and the average surface density and volume densities of the source are given in columns (6)--(8) of Table\,\ref{tab:physical}. 

The velocity dispersion contributed by the thermal motion of \htco~molecules is given by
\begin{equation} \label{eq:sigmath_h2co}
    \sigma_{\ssstyle \rm th,H_2CO} = \sqrt{\frac{k_B T_{\rm kin}}{m_{\rm H_2CO}}},
\end{equation}
where $T_{\rm kin}$ is kinetic temperature derived in Section\,\ref{results:temperature} and $m_{\rm H_2CO}$ is the molecular weight 30 times $m_H$. With $\sigma_{\rm H_2CO}$ deduced from the observed velocity dispersion $\sigma_{\rm obs}$, the non-thermal velocity dispersion is derived as
\begin{equation} \label{eq:sigmant}
    \sigma_{\rm nt} = \sqrt{\sigma^2_{\rm obs}-\Delta_{\rm chan}^2/(2\sqrt{2\ln 2})^2-\sigma^2_{\rm th,H_2CO}},
\end{equation}
where $\Delta_{\rm chan}=1.34$\,\kms~is the channel width. We check that 30 sources with line width $>7$\,\kms~are strongly influenced by outflow wings (see Appendix\,\ref{app:h2co_fit}), and set an averaged observed velocity dispersion of $1.66$\,\kms.

The gas sound speed ($c_s$) or thermal velocity dispersion ($\sigma_{\rm th}$) is given by, 
\begin{equation} \label{eq:sigmath}
    c_s = \sigma_{\rm th} = \sqrt{\frac{k_B T_{\rm kin}}{\mu_{p} m_{\ssstyle \rm H}}},
\end{equation}
where $\mu_{p}=2.37$ for a mean molecular weight per free particle \citep{Kauffmann2008dust}.
The three-dimensional (3D) Mach number is defined as,
\begin{equation} \label{eq:Mach}
    \mathcal{M} = \frac{\sqrt{3}\sigma_{\rm nt}}{c_s}.
\end{equation}
The sound speed, non-thermal velocity dispersion and Mach number are listed in columns (9)--(11) of Table\,\ref{tab:physical}. 

\begin{table*}[!htb]
\centering
\caption{Physical Parameters of ACA 1.3\,mm Continuum Sources \label{tab:physical}}
\setlength{\tabcolsep}{5pt}
\renewcommand{\arraystretch}{1.5} 
\begin{tabular}{ccccccccccc}
\hline
\hline
Field & ID & $T_{\rm kin}$ & $M_{\rm source}$  & $R_{\rm dec}$ & $N^{\rm peak}_{\rm H_2}$ & $\Sigma$ & $n_{\rm H_2}$ & $c_s$ & $\sigma_{\rm nt}$ & $\mathcal{M}$ \\
& ACA & (K) & (\msun) & (pc) & (cm$^{-2}$) & (g\,cm$^{-2}$) & (cm$^{-3}$) & (\kms) & (\kms) & \\
(1) & (2) & (3) & (4) & (5) & (6) & (7) & (8) & (9) & (10) & (11) \\
\hline
I13291-6229 & 1 & 32.8(6.3) & 2.7(1.4) & -- & 1.7(0.7)$\times10^{22}$ & -- & -- & 0.34 & 1 & 3.0 \\
I13291-6229 & 2 & 74.8(1.4) & 8.1(3.5) & 0.054 & 3.3(1.2)$\times10^{22}$ & 0.18(0.08) & 1.8(0.8)$\times10^{5}$ & 0.51 & 1.4 & 2.7 \\
I13291-6229 & 3 & 56.6(0.7) & 3.5(1.5) & 0.032 & 1.8(0.7)$\times10^{22}$ & 0.23(0.10) & 3.7(1.6)$\times10^{5}$ & 0.44 & 0.93 & 2.1 \\
I13291-6229 & 4 & 66.6(1.8) & 2.3(1.0) & -- & 1.1(0.4)$\times10^{22}$ & -- & -- & 0.48 & -- & -- \\
I13291-6249 & 1 & 87.4(3.4) & 281.3(122.0) & 0.3 & 8.9(3.3)$\times10^{22}$ & 0.21(0.09) & 3.6(1.6)$\times10^{4}$ & 0.55 & 1.9 & 3.5 \\
I13295-6152 & 1 & 44.5(0.6) & 26.3(11.3) & 0.12 & 5.4(2.0)$\times10^{22}$ & 0.13(0.05) & 5.6(2.4)$\times10^{4}$ & 0.39 & 0.96 & 2.5 \\
I13471-6120 & 1 & 78.0(0.9) & 119.7(51.6) & 0.048 & 2.2(0.8)$\times10^{23}$ & 3.51(1.51) & 3.8(1.6)$\times10^{6}$ & 0.52 & 1.5 & 2.8 \\
I13484-6100 & 1 & 73.0(14.5) & 119.6(61.4) & 0.1 & 9.9(3.9)$\times10^{22}$ & 0.74(0.38) & 3.7(1.9)$\times10^{5}$ & 0.5 & 2.5 & 5.0 \\
I14013-6105 & 1 & 98.0(2.5) & 83.1(35.9) & 0.11 & 1.2(0.5)$\times10^{23}$ & 0.47(0.20) & 2.2(1.0)$\times10^{5}$ & 0.58 & 1.5 & 2.6 \\
I14050-6056 & 1 & 120.0(6.9) & 18.4(8.0) & 0.11 & 3.5(1.3)$\times10^{22}$ & 0.10(0.04) & 4.9(2.1)$\times10^{4}$ & 0.65 & 1.5 & 2.3 \\
\hline
\end{tabular}
\begin{flushleft}
Field name and the continuum source ID are listed in columns (1)--(2). Kinetic temperature is listed in column (3). Source mass ($M_{\rm source}$), deconvolved size ($R_{\rm dec}$), peak column density ($N^{\rm peak}_{\rm H_2}$), surface density ($\Sigma$), and volume density ($n_{\rm H_2}$) are listed in columns (4)--(8). The uncertainties of these parameters (except for $R_{\rm dec}$) are included in parentheses. Unresolved sources have `--' in columns (5), (7) and (8). Sound speed ($c_s$), non-thermal velocity dispersion ($\sigma_{\rm nt}$), and Mach number ($\mathcal{M}$) are listed in columns (9)--(11). If \htco~fitting fails, then mark with `--' in columns (10)--(11). Only a part of the table is shown and the complete table is available in its entirety in machine-readable form. 
\end{flushleft}
\end{table*}

\section{Discussion} \label{sec:discuss}

\subsection{Nature of ACA Source} \label{discuss:nature}

\begin{figure*}[!t]
\centering
\includegraphics[width=1.0\linewidth]{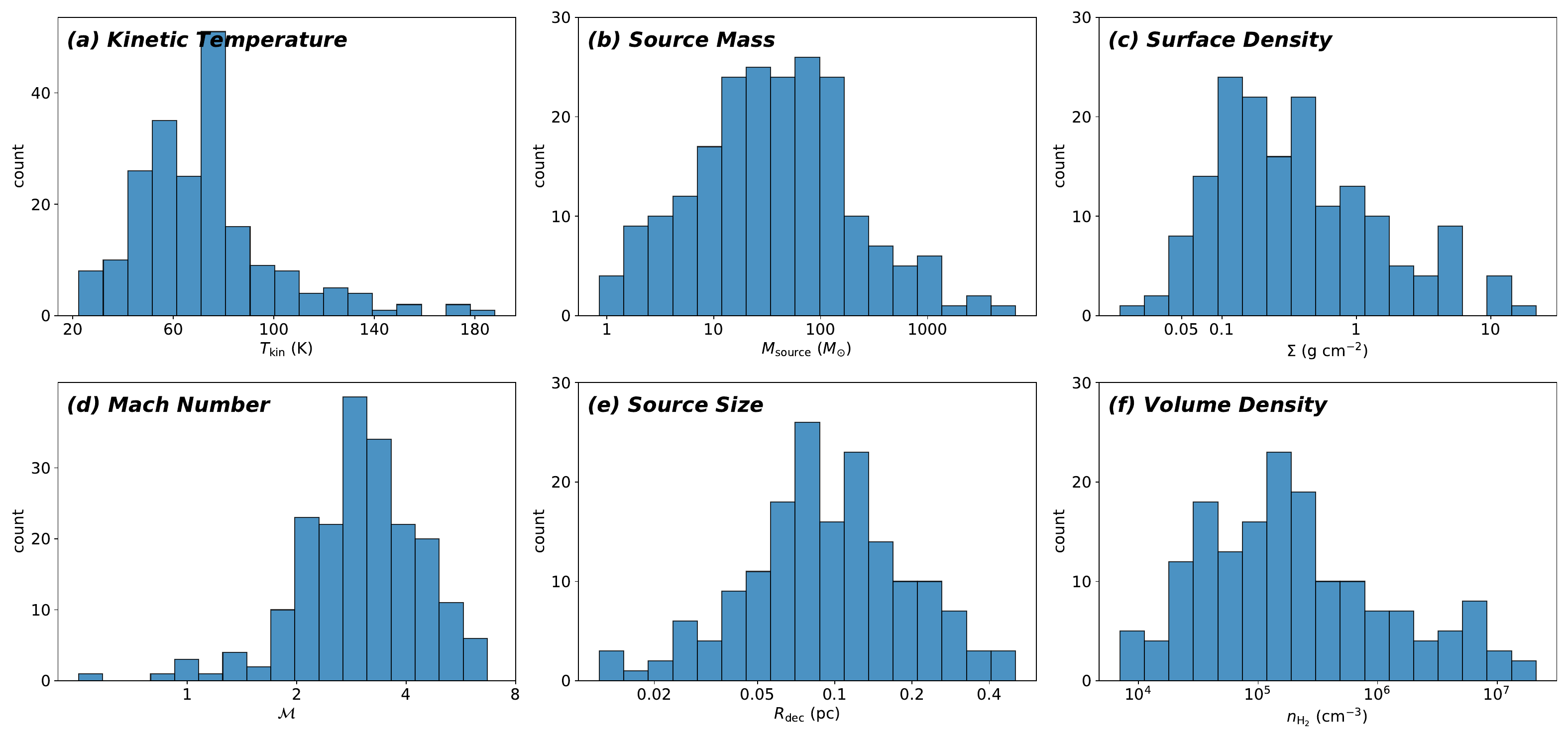}
\caption{Histograms of (a) kinetic temperature $T_{\rm kin}$, (b) source mass $M_{\rm source}$, (c) surface density, (d) Mach number $\mathcal{M}$, (e) source size $R_{\rm dec}$, and (f) volume density $n_{\rm H_2}$. 
\label{fig:stats}}
\end{figure*} 

\subsubsection{Massive star-forming regions with supersonic turbulence} \label{nature:massive}

Six parameters from Table\,\ref{tab:physical}, including kinetic temperature ($T_{\rm kin}$), source mass ($M_{\rm source}$), source size ($R_{\rm dec}$), volume density ($n_{\rm H_2}$), surface density ($\Sigma$) and Mach number ($\mathcal{M}$) are shown in histogram form in Figure\,\ref{fig:stats}. 

In panel (a), $T_{\rm kin}$ ranges from $\sim20$ to $\sim180$\,K, with a mean and median of $\sim$68\,K. More than 97\% sources have a temperature larger than 30\,K, indicating their protostellar nature. With a higher resolution of 2\arcsec, \citet{Qin2022ATOMS-VIII} identified 60 hot molecular cores with gas temperature $>100$\,K, as the heating sources of massive clumps. As shown in Figure\,\ref{fig:temp}, the temperatures of ACA sources are all larger than those of clumps, indicating that the ACA sources are heating their parent massive clumps. Theoretically, a massive protostellar embryo heats and eventually ionizes the gas of its surrounding envelope, creating an \hii~region that develops by expanding within the cloud \citep{Motte2018Review}. Therefore, the embedded heating sources are expected, indicating ACA sources should be heating their parent massive clumps. 

Panel (b) of Figure\,\ref{fig:stats} shows the mass distribution, with 16 and 84 percentile values of 6 and 160\,\msun.  Assuming monolithic collapse and that only a single star forms, 120 ACA sources (58\%) are massive enough to form a massive star ($>8$\,\msun) with a core-to-star efficiency $\sim30$\% \citep{Matzner2000Efficiency,Federrath2012SFR}. However, the ACA sources have fragmented and are forming protoclusters (see Section\,\ref{nature:ensemble}). If assuming that a cluster will form inside the ACA sources and using the empirical relation from star clusters given by \citet{Larson2003IMF}, 
\begin{equation} \label{eq:cluster}
    \left(\frac{m_{\rm max}}{M_\odot}\right) = 1.2\left(\frac{M_{\rm cluster}}{M_{\odot}}\right)^{0.45},
\end{equation}
where $m_{\rm max}$ and $M_{\rm cluster}$ are the maximum mass and the total mass of the stellar cluster, we find that 72 out of 207 sources can form massive stars if only mass \textit{in situ} participates in star formation. 
From panel (c), we observe that nearly all the sources possess surface densities ($\Sigma$) exceeding 0.05\,g\,cm$^{-2}$, which aligns with the empirical threshold for high-mass star formation as suggested by \citet{Urquhart2014ATLASGAL}. A total of 35 ACA sources exceed the more stringent surface density threshold of 1\,g cm$^{-2}$ proposed by \citet{Krumholz2008Threshold}. 
But it's essential to bear in mind that these surface density thresholds can be scale-dependent. For example, if massive clumps have a density profile of $\rho\propto R^{-2}$, then the enclosed mass scales with $M\propto R$, resulting in $\Sigma\propto R^{-1}$. If we adjust for this relation, the surface density threshold for massive star formation should be approximately ten times higher at the scale of ACA sources. But in a turbulent-dominated clump, one argues that cores have a column density comparable to that of the clump as a whole, with only $\Sigma_{\rm core}/\Sigma_{\rm cl}=1.22$. 

Core-scale (0.1\,pc and $n>10^5$\,cm$^{-3}$) infall motions have been statistically studied \citep{WuJ2003Infall,WuY2007Infall,Contreras2018G331,Xu2023JCMTInfall} and filamentary accretion flows are resolved in high-resolution observations
\citep{Peretto2013SDC335,Liu2016NH3,Yuan2018G22,Lu2018Filament,Chen2019G14,Sanhueza2021I18089,Redaelli2022AG14,Xu2023ATOMS-XV,Yang2023G310}. Hence, the identified ACA sources are likely to continue to accumulate mass throughout their evolution, achieving further growth of the core mass and enhancement of the surface density \citep{Liu2023HFS,Xu2024ASSEMBLE}. Using H$^{13}$CO$^+$\,(1--0), \citet{Zhou2022HFS} identified 68 hub-filament systems with clear velocity gradients in the ATOMS survey. In the context of massive cluster formation, these hub-filament structures play a crucial role in supporting gas accretion towards dense cores where massive stars form.

In panel (d), the Mach numbers $\mathcal{M}$ at the scale of the ACA sources are mostly greater than 2. We note that those sources with strong line wings are excluded when analyzing the line widths because of the widening effects of the \htco~outflows as reported in \citep{Izumi2023ASHES-X}. Assuming that turbulence dominates the non-thermal motion, the $\mathcal{M}$ values suggest the prevalence of supersonic turbulence, which aligns with what has been found earlier in several cases \citep[e.g.,][]{Zhang2009G28,Wang2014Snake}. As a result, the supersonic turbulence can suppress thermal Jeans fragmentation \citep[$\sim1$\,\msun;][]{Sanhueza2019ASHES} and enhances mass accretion onto central massive protostars, which is proposed in the model by \citet{McKee2003TC}. 

\subsubsection{Fragments with self-similar gravitational collapse} \label{nature:dense}

\begin{figure*}[!t]
\centering
\includegraphics[width=0.8\linewidth]{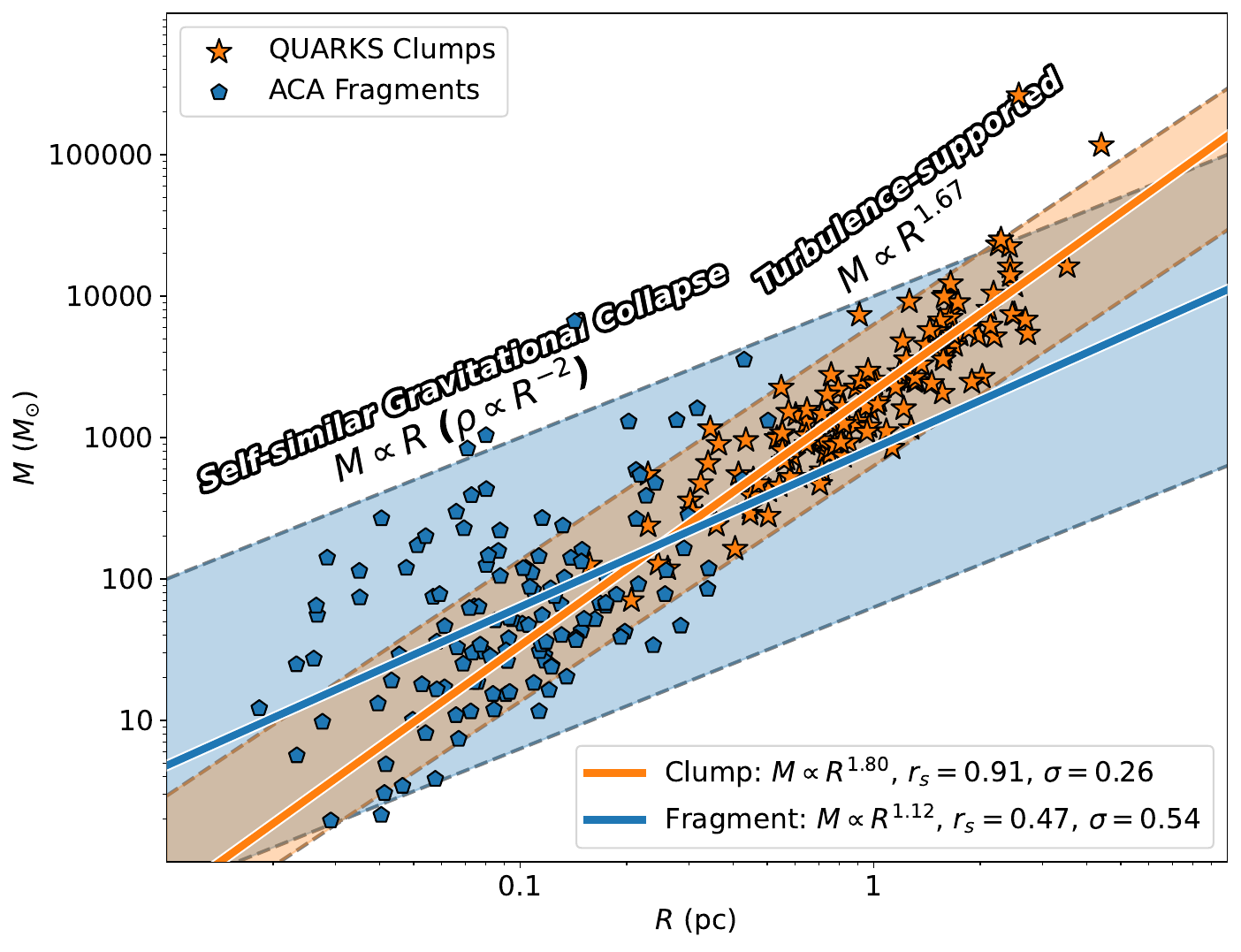}
\caption{Mass versus radius ($M$--$R$) diagram. The QUARKS clumps and the ACA fragments are shown with orange stars and blue pentagons, respectively. Two linear regressions are applied to the QUARKS clumps and the ACA fragments in logarithmic space. The corresponding scaling relations $M\propto R^{1.80}$ and $M\propto R^{1.11}$ are shown with orange and blue solid lines. The Spearman correlation coefficient and 1$\sigma$ scatter are shown in the lower right corner of the figure. The orange and blue shaded regions correspond to the turbulence-dominated ($M\propto R^{2}$) and gravity-dominated ($M\propto R$) regimes with $2\sigma$ data scatters in linear regression fittings. \label{fig:MR}} 
\end{figure*}

According to panels (e) and (f) in Figure\,\ref{fig:stats}, the ACA sources exhibit sizes ranging from 0.02 to 0.4\,pc, volume densities exceeding $10^4$\,cm$^{-3}$, with 107 of them surpassing $10^5$\,cm$^{-3}$. The median volume density of the ACA sources is $1.6\times10^5$\,cm$^{-3}$, which is 20 times larger than that of the clumps. This suggests that these ACA sources are the condensed gas fragments at a subclump scale. Therefore, the ACA sources serve as sub-clump structures and are also called (ACA) fragments in the following discussion. 

We collect the QUARKS clump mass $M_{\rm clump}$ and radius $R_{\rm clump}$ from \citet{Liu2023QUARKS} and plot them (orange stars) with ACA fragments (blue pentagons) in Figure\,\ref{fig:MR}. The source extraction algorithm of the clumps is the same as we do for ACA fragments, so there is no systematic bias due to the methodology in the following discussion. The mass versus radius ($M$--$R$) diagram can be used to study mass concentration at different scales. Sources that shares the same density profile should exhibit scaling relations with the same power law index in the $M$--$R$ diagram. 

We perform a linear regression on the $M$--$R$ diagram for the QUARKS clumps, resulting in a correlation of $M\propto R^{1.8}$ (shown as orange line) with correlation coefficient of 0.91 and $1\sigma$ data scatter of 0.26. Assuming that all the QUARKS clumps share a similar density profile, then the power-law index corresponds to the expectation of the turbulent-support model proposed by \citet{Li2017Isolation}. In their model, energy dissipation rate of external turbulence balances with that of internal virialized turbulence, resulting in a mass concentration relationship described by $M\propto R^{1.67}$. 

The agreement between observations and theory suggests that the QUARKS massive clumps are currently in a transitional phase. In a turbulence-dominated cloud, a shallower density profile following $\rho\propto R^{-1}$ \citep{BT2012} and a steeper mass concentration of $M(<r)\propto R^{2}$ are expected. In contrast, in a dense structure dominated by gravity and the system reaches a quasi-stationary stage \citep{Xu2023ATOMS-XV} after relaxation processes, the density profile usually adheres to $\rho\propto R^{-2}$ \citep{Li2018Selfsimilar}, with the mass concentration described by $M(<R)\propto R$. 

As shown in blue pentagons of Figure\,\ref{fig:MR}, the ACA fragments show a notable deviation from clump (orange color) at scales $\leq0.1$\,pc, suggesting a different mass-radius relationship. These fragments have a higher mass for a given radius compared to the turbulent-supported model. A linear regression applied to the ACA fragments reveals a correlation of $M\propto R^{1.1}$, represented by a blue line, suggesting a power-law index close to 1. This scaling aligns with a density profile of $\rho\propto R^{-2}$, indicative of a scale-free gravitational collapse in a self-similar fashion \citep{Li2018Selfsimilar}. In Section\,\ref{DGF:hierarchy}, self-similarity of ACA fragments will be proposed in another manner. 

The transition of mass concentration from several parsecs to tenths parsec is highly consistent with what has been found by \citet{Peretto2023Decouple}, where star cluster progenitors are reported to be dynamically decoupled from their parent molecular clouds, exhibiting steeper density profiles $\rho\propto R^{-2}$ and flat velocity dispersion profiles $\sigma\propto R^0$, clearly departing from Larson's relations. Similar scale-dependent gas dynamics have been found in several cases of multi-scale studies \citep{Liu2022ATOMS-IX,Saha2022ATOMS-XII}, where gravity-driven gas motion takes over turbulence. In the QUARKS sample, it would be of great interest to investigate the behaviour of $M$--$R$ down to dense core scale as a follow-up work, similar to what has been done in infrared dark clouds \citep[e.g.][]{Li2023ASHES}. 

\subsubsection{Protocluster ensembles} \label{nature:ensemble}

As shown in high-resolution studies by \citet{Xu2024ASSEMBLE} and \citet{Liu2023QUARKS}, the QUARKS ACA 1.3\,mm continuum sources contain protoclusters with a large number of dense cores, which are embedded in the parent ACA sources \citep[see also][]{Zhang2021HMSCs}. To further demonstrate this ubiquity, we performed a spatial cross-match between the ATOMS 3\,mm continuum dense cores by \citet{Liu2021ATOMS-III} and the QUARKS ACA 1.3\,mm continuum sources. 

As a result, we have identified a total of 301 ``QUARKS-ATOMS links'' (links hereafter), as defined by Eq.\ref{eq:ellipse} discussed in Appendix\,\ref{app:crossmatch} and listed them in Table\,\ref{tab:crossmatch}. The remaining 128 ATOMS dense cores, without any associated ACA sources, are referred to as ``field sources''. The presence of field sources can be attributed to the generally higher mass sensitivity of the ATOMS data compared to the QUARKS ACA data. Given a distance of 3\,kpc and dust temperature of 20\,K, typical ATOMS and QUARKS 7-m continuum sensitivities of 0.2 and 15\,mJy give sensitivity limits of 1.5 and 3\,\msun, respectively. Compared to 1.3\,mm, the 3\,mm continuum emission can be contaminated by free-free emission more easily and diffuse 3\,mm emission cannot be seen in the ACA 1.3\,mm continuum images. 

There are 86 ACA sources (42\%) with more than one ATOMS dense cores, and among them, 25 have three or more ATOMS dense cores, indicating the presence of substructures. Consequently, the detected QUARKS ACA sources are likely to be ensembles of protoclusters in nature. However, there are 95 ACA sources associated with single ATOMS dense cores, possibly due to limited resolution. Above all, the analyses at the scale of ACA sources provide a global view of massive protoclusters. 

\subsection{Correlation Between Clumps and Fragments} \label{discuss:corr}

\begin{figure*}[!t]
\centering
\includegraphics[width=1.0\linewidth]{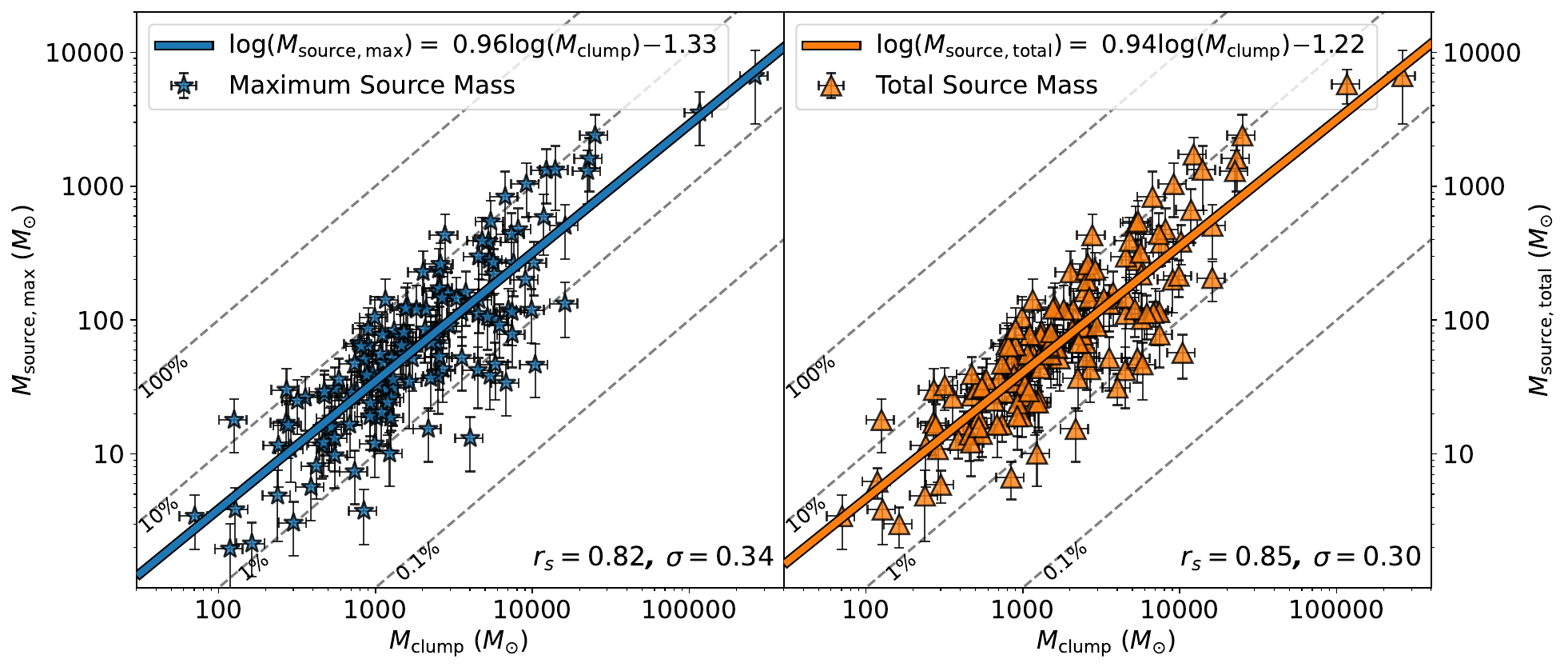}
\caption{Maximum ACA source mass $M_{\rm source,max}$ and total ACA source mass $M_{\rm source,total}$ versus QUARKS clump mass $M_{\rm clump}$ are shown in the left and right panels, respectively. The dashed lines label the cases where $M_{\rm source,max}$/$M_{\rm source,total}$ equals to 0.1, 1, 10, and 100 percent of $M_{\rm clump}$. Linear regressions are performed to fit the data in logarithmic space. The derived scaling relations are: 1) $\log M_{\rm source,max} = 0.96 \log M_{\rm clump} - 1.33$ (blue line), with correlation coefficient of 0.82 and 1$\sigma$ data scatter of 0.34 dex; 2) $\log M_{\rm source,total} = 0.94 \log M_{\rm clump} - 1.22$ (orange line), with correlation coefficient of 0.85 and 1$\sigma$ data scatter of 0.30 dex. \label{fig:Maca_Mclump}} 
\end{figure*}

As discussed in Section\,\ref{nature:dense}, the ACA 1.3\,mm continuum sources trace dense fragments ($n_{\rm H_2}>10^4$\,cm$^{-3}$) within massive clumps. Figure\,\ref{fig:Maca_Mclump} presents maximum mass of ACA sources ($M_{\rm source,max}$) and the total mass of ACA sources ($M_{\rm source,total}$) versus their natal clump mass ($M_{\rm clump}$). These are shown with blue stars and orange triangles, respectively. Comparing the two panels, we find no clear difference between $M_{\rm source,max}$ and $M_{\rm source,total}$ in most cases, because the most massive ACA sources are dominated by mass within clumps. 

Linear regression is used to correlate $M_{\rm source,max}$ with $M_{\rm clump}$. The derived scaling relation is $\log (M_{\rm source,max}/M_{\odot}) = 0.96 (\log M_{\rm clump}/M_{\odot}) - 1.33$ as indicated by blue line, with the correlation coefficient of 0.82 and 1$\sigma$ data scatter of 0.34 dex. To examine the distance effects on the mass correlation, we further perform linear regression in narrower distance bins (see Appendix\,\ref{app:dist}). 

Consistent with our results, \citet{Lin2019SABOCA} find $M_{\rm source,max}\propto M_{\rm clump}^{0.96}$ in the 350\,$\mu$m observations of 204 ATLASGAL clumps. Besides, \citet{Traficante2023SQUALO-I} also find a scaling relation of $M_{\rm source,max}\propto M_{\rm clump}^{1.02}$ in an ALMA survey of 13 massive clumps. However, the quasi-linear correlation could be a result of an evolutionary process, as it may not be evident in the early stages of massive clumps. For example, samples of massive starless clumps have shown a significantly small amount of mass stored in fragments \citep{Sanhueza2019ASHES,Svoboda2019HMSC,Morii2023ASHES}, consistent with the idea that initial fragmentation in massive clumps are Jeans-like and producing low-mass cores. As a clump evolves, the continuous mass accretion feeds the most massive core \citep{Xu2023ATOMS-XV} and the mass correlation builds up \citep{Xu2024ASSEMBLE}. The QUARKS sample predominantly covers mid- and late-stage massive star-forming clumps, characterized by luminosity-to-mass ratio range of 4--460\,$L_{\odot}/M_{\odot}$ and a median value of 35\,$L_{\odot}/M_{\odot}$, so the mass correlation is expected. 

As indicated by orange line in the right panel of Figure\,\ref{fig:Maca_Mclump}, linear regression gives $\log (M_{\rm source,total}/M_\odot) = 0.94 (\log M_{\rm clump}/M_{\odot}) - 1.22$. Intriguingly, the correlation is even tighter with coefficient of 0.85 and 1$\sigma$ data scatter of 0.30 dex, compared to the that of $M_{\rm source,max}$--$M_{\rm clump}$. The tightening correlation indicates that total ACA source (dense gas) mass could be a physical value that is more directly correlated with clump mass. In other words, the $M_{\rm source,max}$--$M_{\rm clump}$ relation could be a combined result of $M_{\rm source,total}$--$M_{\rm clump}$ and a mass function which correlates $M_{\rm source,total}$ with $M_{\rm source,max}$ \citep[e.g.,][]{Bonnell2004Simulation,Weidner2013Cluster}. 

\subsection{Dense Gas Fraction and Its Assembly} \label{discuss:DGF}

Star formation takes place in dense molecular gas. Here we take the ACA 1.3\,mm continuum sources as ``dense gas'' relative to the total clump gas, and define the dense gas fraction (DGF),
\begin{equation}
    \mathrm{DGF} \equiv \frac{M_{\rm dense}}{M_{\rm clump}},
\end{equation}
where $M_{\rm dense}=\sum\limits^{\rm \in clump} M_{\rm source}$ is the total ACA source mass within the clump. \citet{Urquhart2018ATLASGAL} performed the photometry from near- to far-infrared data of these massive clumps, by which the spectral energy distribution (SED) are fitted. By this method, the clump mass es, $M_{\rm clump}$, were derived and are scaled with the updated distances in this work. 

As discussed in Section\,\ref{discuss:corr}, the linearity in Figure\,\ref{fig:Maca_Mclump} indicates a constant DGF within the QUARKS sample, giving a value directly by its intercept of $10^{-1.22}$, i.e., 6\%. In an alternative definition of core formation efficiency (CFE), a highly consistent median value of 6\% is also reported in \citet{Traficante2023SQUALO-I}, evidently consistent with what has been found here. 

Although showing invariance with $M_{\rm clump}$, DGF still has a scatter of 1--10\%. To further explore the origins of scatter, DGF versus clump radius ($R_{\rm clump}$) and luminosity-to-mass ratio ($L/M$) diagrams are explored in Section\,\ref{DGF:hierarchy} and Section\,\ref{DGF:growth}, respectively

\subsubsection{Self-similarity in protocluster formation} \label{DGF:hierarchy}

\begin{figure*}[!t]
\centering
\includegraphics[width=1.0\linewidth]{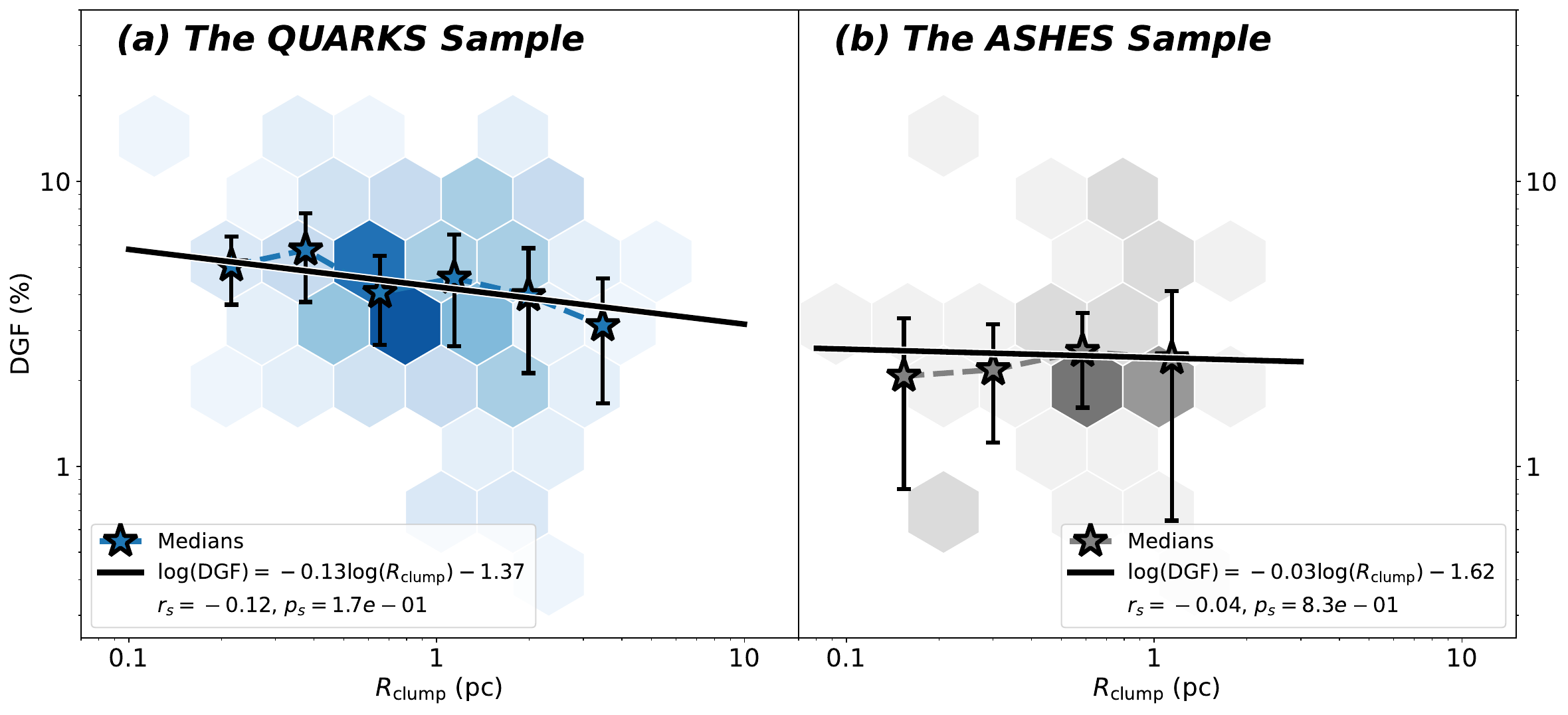}
\caption{Dense gas fraction (DGF) versus clump radius ($R_{\rm clump}$) of (a) the QUARKS sample and (b) the ASHES sample. The hexagons indicate the probability distributions of data points. The colored stars show the median values with errorbars in the $R_{\rm clump}$ bins. 
\label{fig:DGF_Rclump}} 
\end{figure*} 

As shown in the panel (a) of Figure\,\ref{fig:DGF_Rclump}, the blue hexagons indicate the probability distributions of data points in the DGF versus clump radius diagram. The median DGF over $R_{\rm clump}$ bin, as indicated by the blue stars connected by line, shows no discernible systematic variations with $R_{\rm clump}$. Linear regression shows a weak correlation with a Spearman correlation coefficient of -0.12, indicating dense gas mass remains nearly constant relative to the clump mass across different scales from several tenths of parsec to several parsec. It's worth noting that there are some clumps with very low DGF ($<$1\%), indicating that QUARKS field of views are not large enough to cover all the dense gas in clumps with large sizes. However, the coverage-limit effect can be neglected in our sample because the QUARKS pointings are biased to the dense regions according to the ATOMS survey. 

If we consider the ACA sources as what has been observed at various scales, that is, from parsec-scale clumps to sub-parsec-scale cores, then the multi-scale invariant DGF suggests that the gas tends to condense or fragment into dense structure with some constant ratio in a hierarchical system. It implies a self-similar fragmentation or collapsing mode in protocluster formation, as proposed in some case studies \citep[e.g.,][]{Wang2011G28,Wang2014Snake}. More importantly, \citet{Dib2023Protocluster} performed delta-variance spectrum analysis of 15 ALMA-IMF cloud structures \citep{Motte2022ALMA-IMF-1} and discovered a self-similar regime $\lesssim$\,0.03--0.3\,pc. Referring to panel (a) of Figure\,\ref{fig:DGF_Rclump}, one can find it highly consistent with the sizes of the ACA sources, representing the most compact clumps within the protocluster forming clouds \citep{Dib2023Protocluster}. Therefore, our result favors the self-similarity of density structure in massive protoclusters in a dependent way. 

We also retrieved 39 ACA 1.3\,mm continuum images from the ``ALMA Survey of 70\,$\mu$m Dark High-mass Clumps in Early Stages'' \citep[ASHES hereafter][]{Sanhueza2019ASHES,Morii2023ASHES}. It is worth noting that the ASHES observations adopted a mosaic mode with larger fields of view than the QUARKS. To maintain consistency, we cropped the ASHES continuum images to the same size as the QUARKS. After adopting the same source extraction algorithm and mass calculation, we derive the DGF of 39 ASHES clumps which are shown with gray hexagons in the panel (b) of Figure\,\ref{fig:DGF_Rclump}. The gray stars indicate the median values in corresponding parameter bins. Similarly, the ASHES sample also shows an invariance of DGF with size. Therefore, the self-similarity works in both early and late stages of the evolution of massive protoclusters. 

\subsubsection{Dense gas grows with evolution} \label{DGF:growth}

\begin{figure*}
\centering
\includegraphics[width=1.0\linewidth]{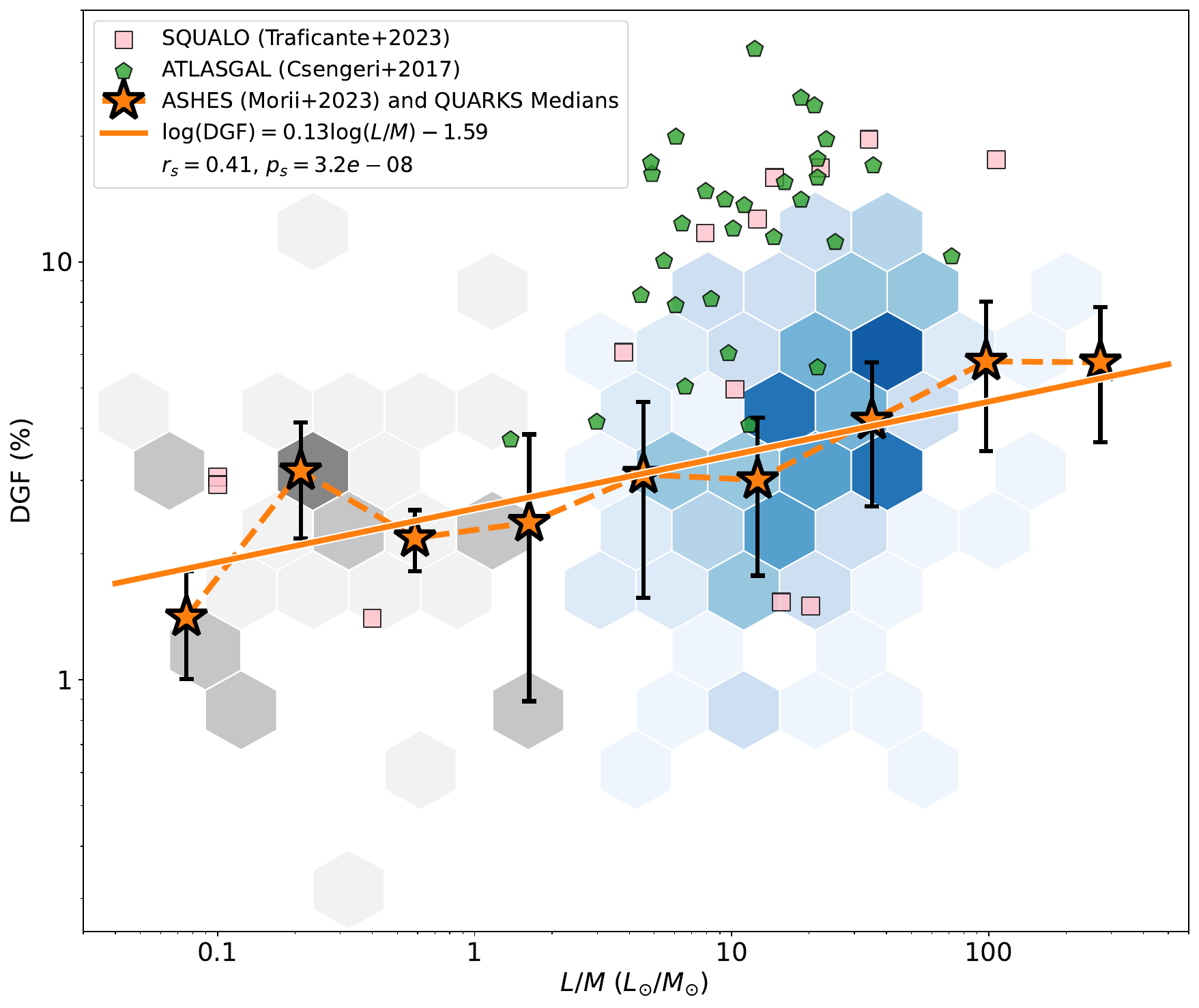}
\caption{Dense gas fraction (DGF) versus clump luminosity-to-mass ratio ($L/M$), indicative of clump evolutionary stage. Blue and gray hexagons show the probability distribution for the QUARKS and ASHES samples \citep{Morii2023ASHES}, with orange stars showing the median values in the parameter bins. The orange stars connected by dashed lines show the median DGF values of the combined sample QUARKS + ASHES in the $L/M$ bins from 0.04 to 400\,$L_\odot/M_\odot$. Linear regression is adopted for the combined sample, and the orange dashed line, $\log({\rm DGF}) = 0.13\log{(L/M)}-1.59$, illustrates an increasing trend of DGF with $L/M$. This relationship shows a Spearman correlation coefficient of 0.41 and a p-value of $3.2\times10^{-8}$. The pink squares points are retrieved from the SQUALO project \citep{Traficante2023SQUALO-I}, and green pentagons are from the ALMA survey of massive cluster progenitors in ATLASGAL \citep{Csengeri2017ACA}. 
\label{fig:DGF_LtoM}}
\end{figure*}

In Figure\,\ref{fig:DGF_Rclump}, an intriguing feature is a systematically one-time larger DGF in the QUARKS than that in the ASHES clumps, and the DGF difference seems to be invariant with the clump size. Therefore, we refer to the evolutionary explanation of the DGF difference between two samples. Theoretically, a high-mass star grows in mass during its formation process, so its luminosity should increase. The ratio of bolometric luminosity to envelope mass $L/M$, can thus be used to indicate the evolutionary stage \citep{Sridharan2002HMPOs,Elia2017HiGAL}. In Figure\,\ref{fig:DGF_LtoM}, the DGF is plotted against $L/M$, where the gray and blue hexagons represent the probability distribution of data points of the ASHES and the QUARKS sample, respectively. The orange stars connected by dashed show the DGF median values of the QUARKS+ASHES combined sample in the $L/M$ bins from 0.04 to 400\,$L_\odot/M_\odot$, spanning four orders of magnitude. To demonstrate the evolutionary trend, a linear regression is performed on the data points from the combined sample, resulting in a correlation of $\log({\rm DGF}) = 0.13\log{(L/M)}-1.59$, depicted by the orange solid line. The correlation yields a Spearman correlation coefficient of $R_s=0.41$ and an associated $p$ value of $2.9\times10^{-8}$, indicating an evident increase in the mass fraction of the dense part of the clump as it evolves. 

We incorporate the DGF results from an ACA 0.87\,mm survey \citep[C17 hereafter]{Csengeri2017ACA} and the "Star formation in QUiescent And Luminous Objects" (SQUALO) project \citep[hereafter T23]{Traficante2023SQUALO-I}, contributing 30 green and 13 orange data points in Figure\,\ref{fig:DGF_LtoM}. To keep consistency, all the clump masses are retrieved from \citet{Urquhart2018ATLASGAL} in the following discussion. While the $L/M$ of the \citetalias{Csengeri2017ACA} sample spans a similar range to ours, the DGF is systematically larger. This discrepancy arises because \citetalias{Csengeri2017ACA} adopted a constant temperature $T_{\rm dust}=25$\,K throughout the sample, which is systematically lower than what we used in Eq.\,\ref{eq:mass}, resulting in a higher mass. \citetalias{Traficante2023SQUALO-I} improved this method by categorizing the sample into three evolutionary bins and estimating temperatures of 20, 30, and 40\,K, respectively. Despite the limited sample size, the wide range of $L_\odot/M_\odot$ in \citetalias{Traficante2023SQUALO-I} indicates an increasing trend similar to the results of the QUARKS. As a result, two independent data sets mutually verify a dense mass growth in massive star-forming clumps as they evolve. 

The methods in both \citetalias{Csengeri2017ACA} and \citetalias{Traficante2023SQUALO-I} give a lower temperature compared to ours. According to Equation\,\ref{eq:mass}, a lower temperature leads to a higher mass for a given flux. Therefore, we adopt the above temperature estimation methods to our sample, the resulting DGF will be even larger and the increasing trend will become more significant. 

The mass growth of dense cores is widely found during the evolution of massive star-forming clumps \citep{Anderson2021HFS,Traficante2023SQUALO-I,Liu2023HFS,Li2023ASHES,Xu2024ASSEMBLE}. Very recently, the ASHES team has found an increase in the mass dynamic range with respect to the protostellar core fraction, serving as a proxy for evolutionary stages \citep{Morii2024}. Our work, encompassing a significantly larger sample with a broad range of evolutionary stages, substantiates the continuous growth of dense mass over time. Overall, recent ALMA studies portray a dynamic scenario wherein dense gas accumulates throughout the evolutionary process. 

We note that the evolution-dependent DGF cannot explain the total scatter. The wide range of DGFs among the QUARKS samples can also arise from the dynamic balance between gas depletion/stellar feedback and gas infall. Specifically, gas infall and accretion processes function to concentrate the dense gas, while star formation depletes the dense gas. Stellar feedback mechanisms, such as winds and outflows, play a dual role by releasing gas back to the parent clump and preventing further gas accretion. But for the surrounding embedded dense gas structures, their kinematic properties may be less influenced by feedback from the most evolved stars in clumps \citep{Zhou2023Feedback}. At any rate, a systematic investigations into gas kinematics and energetics have the potential to unveil dynamic effects on DGF and elucidate the origin of the scatter in DGF values. 

\subsection{Limited Fragmentation} \label{discuss:fragmentation}

ACA sources are fragments from the clump scale. The facts that the mean number of fragments per clump is $\bar N_{\rm frag}\sim1.5$ and that 93 clumps have only one fragment, both suggest limited fragmentation, which is consistent with what has been found in another ACA survey by \citetalias{Csengeri2017ACA}. 

\citetalias{Csengeri2017ACA} discussed the possibility that global collapse at the clump scale could explain the excessive mass of the subclump reservoir \citep{Schneider2010DR21,Peretto2013SDC335,Xu2023ATOMS-XV}. This concept implies that the entire clump is involved in a dynamic process, wherein fragments and low-density gas experience global collapse. Equilibrium may never be reached at subclump scales, which aligns with the limited fragmentation observed. \citetalias{Csengeri2017ACA} observed that the majority of the clumps are likely not in virial equilibrium, suggesting collapse on the clump scale. Furthermore, continuous mass accretion beyond the clump to the core-scale feed could fuel the formation of the protocluster \citep{Avison2021SDC335,Xu2023ATOMS-XV,Yang2023G310,Xu2023JCMTInfall}. In this scenario, an increase in the number of fragments with time and a Jeans-like fragmentation to develop in more evolved stages are expected \citep{Palau2015Fragmentation}, and appear to be in conflict with our observed evolution-independent limited fragmentation. However, the conflict can be reconciled, because the QUARKS ACA observations have limited mass sensitivity of $\gtrsim3$\,\msun, at a temperature of 20\,K and distance of 4\,kpc and limited spatial resolution of $\gtrsim0.1$\,pc. Therefore, those small low-mass fragments cannot be effectively resolved even if they exist. \citet{Liu2023QUARKS} show an example of Sgr B2(M) in the QUARKS high-resolution ($\sim0.3\arcsec$) data. Compared to the only detection in the ACA data in I17441-2822, \citet{Liu2023QUARKS} identified more than 30 spatially associated cores with the ACA 1.3\,mm source and up to 93 cores in the entire cluster of cores. 

Nevertheless, we propose another possibility where ACA fragments are the products of turbulent Jeans fragmentation from the natal clump. Although the thermal Jeans mass in massive clumps is as low as several~\msun, at clump density of $\bar n_{\rm cl}\sim5\times10^{4-5}$\,cm$^{-3}$ and kinetic temperature of $T_{\rm kin}=20\,K$, turbulent Jeans fragmentation favors more massive fragment, larger Jeans length, and therefore less fragments in a clump.

\section{Summary} \label{sec:sum}

The QUARKS survey, standing for `Querying Underlying mechanisms of massive star formation with ALMA-Resolved gas Kinematics and Structures', is observing 139 massive gas clumps at ALMA Band 6 ($\lambda\sim$ 1.3 mm). This paper introduces the Atacama Compact Array (ACA) 7-m data of the QUARKS survey, describing the ACA observations and data reduction. Combining multi-wavelength data, we provide the first edition of QUARKS atlas, offering insights into the multiscale and multiphase interstellar medium (ISM) in high-mass star formation. 

Leveraging the QUARKS ACA data, we construct the ACA 1.3\,mm continuum source catalog with 207 sources. At least one source and up to five sources are found in one clump. Three source-averaged formaldehyde transition lines p-\htco\,(3--2) are fitted using non-LTE radiative transfer model, to obtain the gas kinetic temperature and line width. Based on the geometric and flux measurements of the ACA sources, and assuming that gas temperature equals the dust temperature, physical parameters including mass and surface/volume densities are derived. The thermal and non-thermal dispersion, as well as the Mach number, are also calculated from fitted p-\htco\,(3--2) line width. 

Statistically speaking, the nature of ACA sources is massive gravity-dominated fragments with at the subclump scale, with supersonic turbulence and possibly embedded star-forming protocluster. A quasi-linear correlation between clump mass and ACA source mass is found, which can be explained by the dynamic coevolution between clump and core in a late stage. The dense gas fraction (DGF) is defined as total ACA source mass over the clump mass, and is found to be about 6\%, although with a wide scatter of 1--10\%. If we consider the massive clump sample as what has been observed at various scales, then the size-independent DGF indicates that the gas conversion efficiency at each scale level remains constant in a hierarchical system, implying a self-similar fragmentation or collapsing mode in protocluster formation. With the data across four orders of magnitude of luminosity-to-mass ratio, we find a significantly increasing trend of DGF with clump evolution. The fragmentation on the subclump scale is limited and the reasons could be that equilibrium may never be reached at subclump scale and massive clumps are undergoing a global collapse. 

\begin{acknowledgements}

We thank the anonymous referee for helpful comments.
This work has been supported by the National Science Foundation of China (12033005), the National Key R\&D Program of China (No. 2022YFA1603102), the China Manned Space Project (CMS-CSST-2021-A09, CMS-CSST-2021-B06), and the China-Chile Joint Research Fund (CCJRF No. 2211). CCJRF is provided by Chinese Academy of Sciences South America Center for Astronomy (CASSACA) and established by National Astronomical Observatories, Chinese Academy of Sciences (NAOC) and Chilean Astronomy Society (SOCHIAS) to support China-Chile collaborations in astronomy. We acknowledge support from the Tianchi Talent Program of Xinjiang Uygur Autonomous Region.
This research was carried out in part at the Jet Propulsion Laboratory, California Institute of Technology, under a contract with the National Aeronautics and Space Administration (80NM0018D0004).
PS was partially supported by a Grant-in-Aid for Scientific Research (KAKENHI Number JP22H01271 and JP23H01221) of JSPS. 
AS, DM, GG, and LB gratefully acknowledge support by ANID Basal project FB210003.
MJ acknowledges the support of the Research Council of Finland Grant No. 348342. 
K.M is supported by a Grants-in-Aid for the the JSPS Fellows (KAKENHI Number 22J21529).
Data analysis was in part carried out on the Multi-wavelength Data Analysis System operated by the Astronomy Data Center (ADC), National Astronomical Observatory of Japan. 

This paper uses the following ALMA data: ADS/JAO.ALMA\#2019.1.00685.S and 2021.1.00095.S. ALMA is a partnership of ESO (representing its member states), NSF (USA) and NINS (Japan), together with NRC (Canada), MOST and ASIAA (Taiwan), and KASI (Republic of Korea), in cooperation with the Republic of Chile. The Joint ALMA Observatory is operated by ESO, AUI/NRAO, and NAOJ. 
The MeerKAT telescope is operated by the South African Radio Astronomy Observatory, which is a facility of the National Research Foundation, an agency of the Department of Science and Innovation.
\textit{Software}. This research uses \textsc{astropy}, a community-developed core Python package for Astronomy \citep{Astropy2013,Astropy2018,Astropy2022}. This research has used the program \textsc{SExtractor}, which builds a catalog of objects from an astronomical image \citep{Bertin1996SExtractor}. This research has used Python-based package \textsc{pyspeckit} to fit spectral lines \citep{Ginsburg2011Pyspeckit,Ginsburg2022Pyspeckit}. This research has used \textsc{RADEX}, a computer program for fast non-LTE analysis of interstellar line spectra \citep{van2007RADEX}. 

\end{acknowledgements}

\clearpage
\appendix

\section{Source Extraction} \label{app:sextractor}

For each field, we initially generated the background and rms map from the original continuum map without applying primary beam correction (\texttt{unpbcor}). Within each unit of boxes, whose size is equivalent to the maximum recoverable scale, we performed iterative clipping of the local background histogram until convergence was achieved at $\pm3\sigma$ around its median \citep{Bertin1996SExtractor}. The background subtracted continuum map, together with the corresponding rms map, was then used as input for \textsc{SExtractor}. Before the program runs, $\texttt{nthresh}=4$ is set to mask the low SNR (\texttt{nthresh}$\times$local rms) pixels. During the extraction procedure, the deblending parameters, that is, the number of thresholds for the deblending, and the deblending contrast are set $\texttt{deblend\_nthresh}=512$ and $\texttt{deblend\_cont}=10^{-5}$. To ensure that the extraction focused on genuine sources rather than spurious features or cleaning artifacts, we set the parameter controlling the minimum pixel count for a source, \texttt{minarea}, to match the effective beam size. 

Due to the Gaussian-like primary beam response, the real flux of source should be corrected by primary beam correction (\texttt{pbcor}). So, the \texttt{pb} map is interpolated to the barycenter and at the intensity peak of the source, by which the measured integrated flux and peak intensity are divided, respectively.

\section{Non-LTE Model Fitting of \texorpdfstring{H$_2$CO}{H2CO}} \label{app:h2co_fit}

Following the method introduced in \citet{Ginsburg2016CMZTemp}, we used \textsc{RADEX} \citep{van2007RADEX} to create model grids for the p-\htco~molecular lines over 100 densities of $n=10^{2.5}$--$10^{7}$\,cm$^{-3}$, 100 \htco~column densities of $N$(\htco)$=10^{11}$--$10^{15}$\,cm$^{-2}$, and 50 kinetic temperatures of $T_{\rm kin}=10$--$350$\,K, with a fixed assumed line gradient of 5\,\kms~pc$^{-1}$. In the grid modeling, the collision rates were taken from \citet{Wiesenfeld2013H2CO} and calculated for temperatures in the range from 10 to 300\,K including energy levels up to about 200\,cm$^{-1}$ for collisions with H$_2$. Based on the preconstructed non-LTE model grids, the line fitting is performed. The five parameters are kinetic temperature ($T_{\rm kin}$), formaldehyde column density ($\log N(\rm H_2CO)$), hydrogen molecule volume density ($\log n(\rm H_2)$), centroid velocity ($v_{\rm lsr}$) and line width (FWHM). For all the sources, only one velocity component is considered. 

\begin{figure*}[!htb]
\centering
\includegraphics[width=\linewidth]{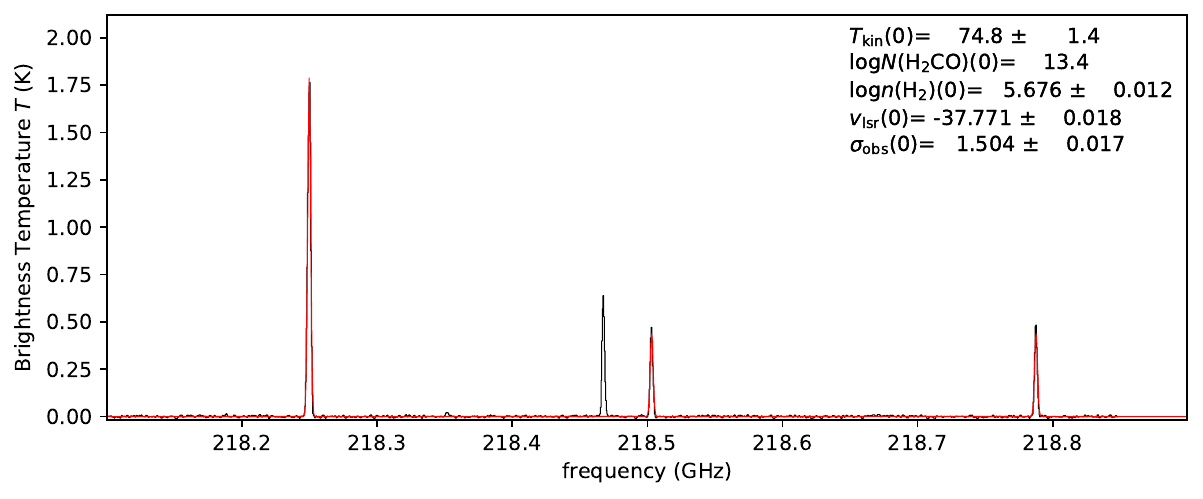}
\caption{As an example, the spectral line data and the fitting model of I13291-6229\_ACA2 are shown in red and black color, respectively. The fitted parameters kinetic temperature $T_{\rm kin}$ (K), formaldehyde column density $\log N(\rm H_2CO)$ (cm$^{-2}$), hydrogen molecule volume density $\log n(\rm H_2)$ (cm$^{-3}$), centroid velocity $v_{\rm lsr}$ (\kms) and FWHM line width (\kms) are shown on the top right. The unfitted line between 218.4\,GHz and 218.5\,GHz is CH$_3$OH $J_K=4_{2,2}-3_{1,2}$.} \label{fig:fitexample}
\end{figure*}

An example (I13291-6229\_ACA2) of spectral line data and the fitting model is shown in Figure\,\ref{fig:fitexample}, where $T_{\rm kin}$ is estimated to be 53($\pm11$)\,K. If the fitting fails (for seven spectra), then the $T_{\rm kin}$ is set to be equal to clump-averaged dust temperature $T_{\rm dust,clump}$ which is retrieved from \citet{Urquhart2018ATLASGAL}. The kinetic temperatures for 207 ACA sources are then listed in column (3) of Table\,\ref{tab:physical}.

\subsection{Validation}

The volume density of the collisional partner, which is molecular hydrogen (H$_2$) in our case, serves as one of the non-LTE model parameters. Assuming a dust emission model and a good mix of dust and gas, it is possible to determine the source-averaged volume density of H$_2$. This value is listed in column (8) of Table\,\ref{tab:physical} independently.

As depicted in panel (f) of Figure\,\ref{fig:stats}, the density of ACA sources predominantly falls within the range of $10^{4}$ to $10^{6}$\,cm$^{-3}$, which is well-suited for \htco~triplet model fitting. When the density surpasses $10^{6}$\,cm$^{-3}$, the line ratio becomes less sensitive to kinetic temperature, primarily due to the effects of radiative trapping, as shown in Figure 6 of \citet{Ao2013CMZTemp}. As we revisit and scrutinize the input parameters, we ensure the validity of our non-LTE model fitting for most of the QUARKS ACA sources. 

\begin{figure*}[!htb]
\centering
\includegraphics[width=0.8\linewidth]{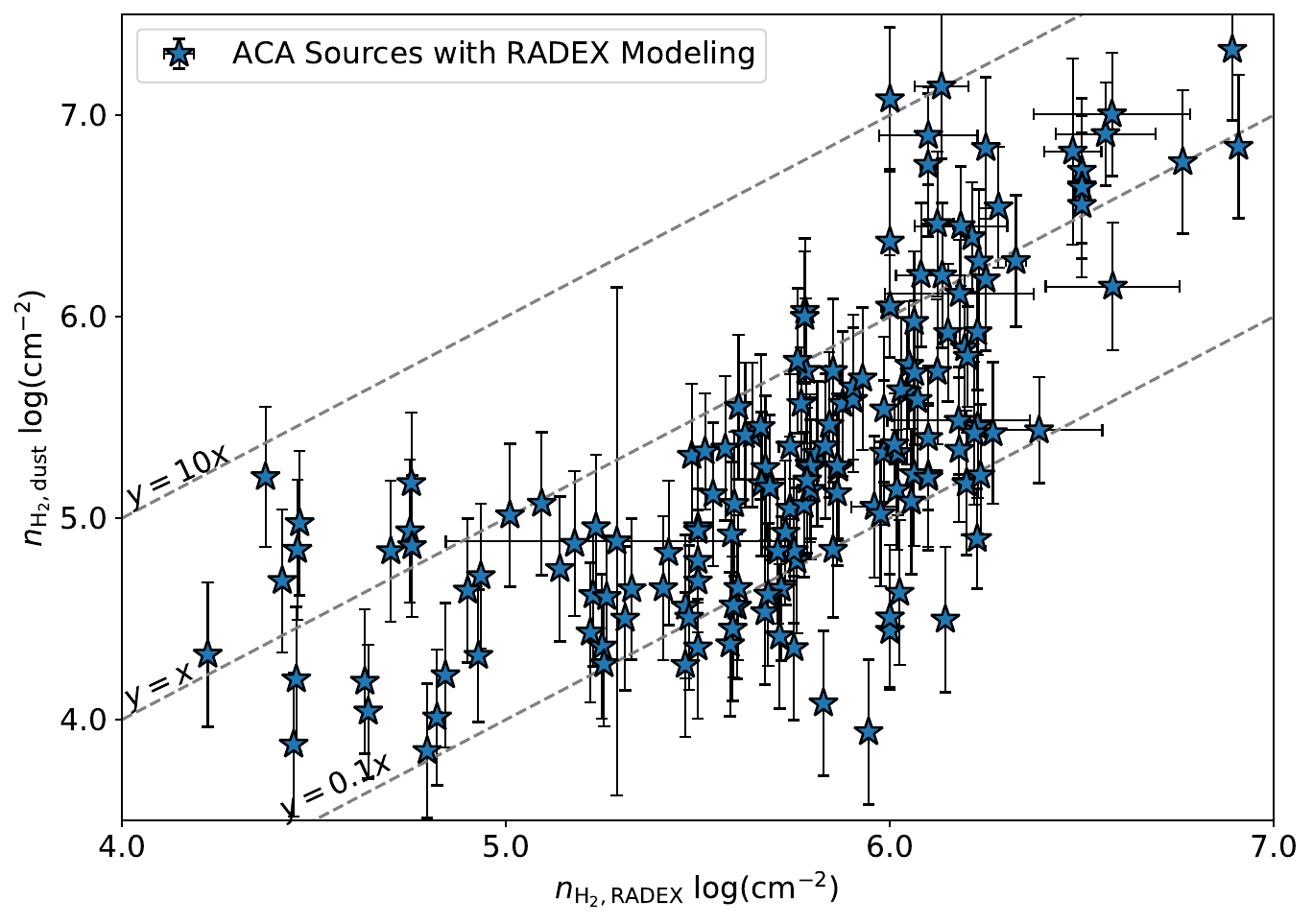}
\caption{Volume density $n_{\rm H_2}$ derived from RADEX line modeling versus that derived from dust emission in a sample of ACA sources with RADEX modeling. The black dashed lines mark $n_{\rm H_2,dust}=0.1/1/10\times n_{\rm H_2,RADEX}$. \label{fig:twodensity}}
\end{figure*}

In Figure\,\ref{fig:twodensity}, we compare the volume density derived from RADEX line modeling $n_{\rm H_2,RADEX}$ and that derived from dust emission $n_{\rm H_2,dust}$ (Section\,\ref{results:physics}). The data points follow a bulk increasing trend although with a large dispersion, further justifying the self-consistency of the temperature estimation. Besides, we note that $n_{\rm H_2,RADEX}$ is systematically higher than $n_{\rm H_2,dust}$, as depicted by black dashed lines. This can be explained by the difference in spatial distribution or the sizes that the \htco/dust trace. If the size that dust traces is higher than that \htco~traces, then $n_{\rm H_2,dust}$ should be naturally lower than 
$n_{\rm H_2,RADEX}$. Therefore, QUARKS high-resolution data should be essential to resolving and understanding internal density structure. 

\subsection{Caveats}

The temperature estimation using the \htco~triplet assumes that the \htco~emission mainly traces dense gas, which is not the case in a sample of 12 infrared dark clouds \citep{Izumi2023ASHES-X}. These authors argued that \htco~emission is mainly sensitive to low-velocity outflow components rather than to quiescent gas expected in the early phases of star formation. So one must be keep in mind the fidelity of the temperature estimated from \htco~triplets depends on how well \htco~emission traces the gas in dense cores. 

The non-LTE line modeling can be influenced by the line profile, including the wings of the line and self-absorption. For example, line wings can make least-square method fall into a false local minimum. There are six cases of severe self-absorption, for which we only assign them $T_{\rm dust,clump}$. If more than one velocity component is associated with an ACA source, the line width will be overestimated. 

\section{QUARKS-ATOMS Link} \label{app:crossmatch}

Elliptical mask is defined by the geometrical measurement of the ACA 1.3\,mm continuum source,
\begin{equation} \label{eq:ellipse}
\begin{split}
    & \frac{\left[(x-x_Q)\cos\phi+(y-y_Q)\sin\phi\right]^2}{A^2} \\
    & + \frac{\left[(x-x_Q)\sin\phi-(y-y_Q)\cos\phi\right]^2}{B^2} \leq 1,
\end{split}
\end{equation}
where ${x,y}$ and ${x_Q,y_Q}$ are respectively the coordinates of the ATOMS and QUARKS sources, $A$, $B$, and $\phi$ are the major axis, minor axis, and the position angle of the QUARKS source. All parameters can be found in Table\,\ref{tab:measurements} and ATOMS Table\,\ref{tab:crossmatch}. A QUARKS-ATOMS source link (link hereafter) is established when Eq.\,\ref{eq:ellipse} is satisfied. 

We list the link, as introduced in Section\,\ref{discuss:nature}, between ATOMS 3\,mm dense cores and QUARKS 1.3\,mm sources in Table\,\ref{tab:crossmatch}. The field name and the ATOMS dense core ID are listed in columns (1)--(2). The Galactic name of dense core, inherited from \citet{Liu2021ATOMS-III}, is listed in column (3). The ICRS coordinates of the barycenter are listed in columns (4)--(5). The associated QUARKS source ID is listed in column (6). If no associated QUARKS source, then the ATOMS sources are referred as ``field sources'', which are marked by ``0''. The angular size ($L_{d}$) and position angle ($L_{\rm PA}$) of the link are listed in columns (7)--(8). The link fidelity, defined as how close the ATOMS source is to the QUARKS source, with values from 0 (unreliable) to 1 (reliable), is listed in column (9). 

\begin{table*}[!thb]
\centering
\caption{ATOMS 3\,mm Dense Cores Linked to QUARKS 1.3\,mm ACA Sources \label{tab:crossmatch}.}
\renewcommand{\arraystretch}{1.5} 
\begin{tabular}{ccccccccc}
\hline
\hline
\multirow{2}{*}{Field} & \multirow{2}{*}{\makecell[c]{ATOMS \\ ID}} & \multirow{2}{*}{Galactic Name} & \multicolumn{2}{c}{Equatorial Coordinates} & \multirow{2}{*}{\makecell[c]{QUARKS \\ ID}} & $L_d$ & $L_{\rm PA}$ & Fidelity \\
\cmidrule(r){4-5} 
& & & RA (ICRS) & DEC (ICRS) & & (\arcsec) & (\degree) & \\
(1) & (2) & (3) & (4) & (5) & (6) & (7) & (8) & (9) \\
\hline
I08303-4303 & 1 & G261.6444-02.0876 & 08:32:09.0 & -43:13:42.9 & 1 & 4.5 & 41.6 & 0.69 \\
I08303-4303 & 2 & G261.6444-02.0890 & 08:32:08.6 & -43:13:45.8 & 1 & 1.2 & 108.5 & 0.99 \\
I08303-4303 & 3 & G261.6446-02.0899 & 08:32:08.4 & -43:13:48.3 & 1 & 3.9 & 57.0 & 0.8 \\
I08448-4343 & 1 & G263.7745-00.4266 & 08:46:35.0 & -43:54:23.8 & 1 & 1.4 & 28.5 & 0.93 \\
I08448-4343 & 2 & G263.7756-00.4281 & 08:46:34.9 & -43:54:30.3 & 2 & 3.5 & 23.7 & 0.57 \\
I08448-4343 & 3 & G263.7756-00.4291 & 08:46:34.6 & -43:54:32.6 & 2 & 1.7 & 123.7 & 0.94 \\
I08448-4343 & 4 & G263.7743-00.4317 & 08:46:33.7 & -43:54:34.8 & 3 & 3.5 & 59.9 & 0.77 \\
I08448-4343 & 5 & G263.7737-00.4326 & 08:46:33.3 & -43:54:35.1 & 3 & 1.7 & 150.5 & 0.91 \\
I08448-4343 & 6 & G263.7712-00.4363 & 08:46:31.8 & -43:54:36.4 & 4 & 7.6 & 95.0 & 0.59 \\
I08448-4343 & 7 & G263.7723-00.4350 & 08:46:32.4 & -43:54:36.6 & 4 & 1.6 & 108.5 & 0.97 \\
I08448-4343 & 8 & G263.7700-00.4379 & 08:46:31.1 & -43:54:36.7 & 0 & -- & -- & -- \\
I08448-4343 & 9 & G263.7744-00.4328 & 08:46:33.4 & -43:54:37.5 & 3 & 1.0 & 168.4 & 0.96 \\
I08448-4343 & 10 & G263.7766-00.4309 & 08:46:34.3 & -43:54:39.4 & 0 & -- & -- & -- \\
I08448-4343 & 11 & G263.7724-00.4366 & 08:46:32.0 & -43:54:40.5 & 4 & 6.7 & 59.6 & 0.19 \\
I08448-4343 & 12 & G263.7729-00.4364 & 08:46:32.1 & -43:54:41.4 & 0 & -- & -- & -- \\
I08448-4343 & 13 & G263.7776-00.4332 & 08:46:34.0 & -43:54:47.4 & 0 & -- & -- & -- \\
\hline
\end{tabular}
\begin{flushleft}
The field name and the ATOMS dense core ID are listed in columns (1)--(2). The Galactic name of dense core is listed in column (3). The ICRS coordinates of the barycenter are listed in columns (4)--(5). The associated QUARKS source ID is listed in column (6). If there is no associated QUARKS source, then ``0''. The angular size ($L_{d}$) and position angle ($L_{\rm PA}$) of the link are listed in columns (7)--(8). The fidelity of the link is listed in column (9). The table is available in its entirety in machine-readable form. 
\end{flushleft}
\end{table*}

Note that there are some ACA sources that lack associated ATOMS 3\,mm sources, which can be categorized into two primary scenarios. In one scenario, the ATOMS dense core catalog prioritizes the high fidelity of genuinely dense cores, but this can lead to a higher false negative rate. Consequently, some QUARKS ACA sources with non-spherical morphologies may be missed in the ATOMS dense core catalog. An example of this is seen in sources I16132-5039\_ACA1 and I16132-5039\_ACA2, which are elongated sources without any associated ATOMS dense cores. In the other scenario, QUARKS data, particularly in certain fields, exhibit greater sensitivity than ATOMS. This higher sensitivity enables the detection of fainter structures. For example, I17269-3312\_ACA3 and I17269-3312\_ACA4, two relatively faint sources in the I17269-3312 field, are marginally seen in the ATOMS data and are not included in the catalog. 

\section{Distance Effects on Mass Correlation} \label{app:dist}

The calculations of the clump and the ACA source masses depend on distance by $M\propto d^2$. Therefore, it is crucial to examine the impact of distance on the mass correlation discussed in Section\,\ref{discuss:corr}. To explore this, we categorize the QUARKS clumps into three groups based on their distances: $d\leq3$\,kpc (Near), $3<d\leq6$\,kpc (Mid), and $d>6$\,kpc (Far). The selection of distance bins is solely based on achieving a similar sample size in each group. 

We perform linear regression on the three groups individually and obtain similar quasi-linear correlations as observed in the total sample. The correlation coefficients for the three fittings are substantial, ranging from 0.72 to 0.74, and the $1\sigma$ scatters fall between 0.25 and 0.35. Both metrics indicate a significant correlation between $M_{\rm clump}$ and $M_{\rm source,total}$. Additionally, we observe wide mass range for narrow color (i.e., distance) range, suggesting that even narrower distance bins would still result in a relatively strong mass correlation. In summary, we assert a weak distance effect on the mass correlation in the QUARKS sample. 

\begin{figure*}[!ht]
    \includegraphics[width=1.0\linewidth]{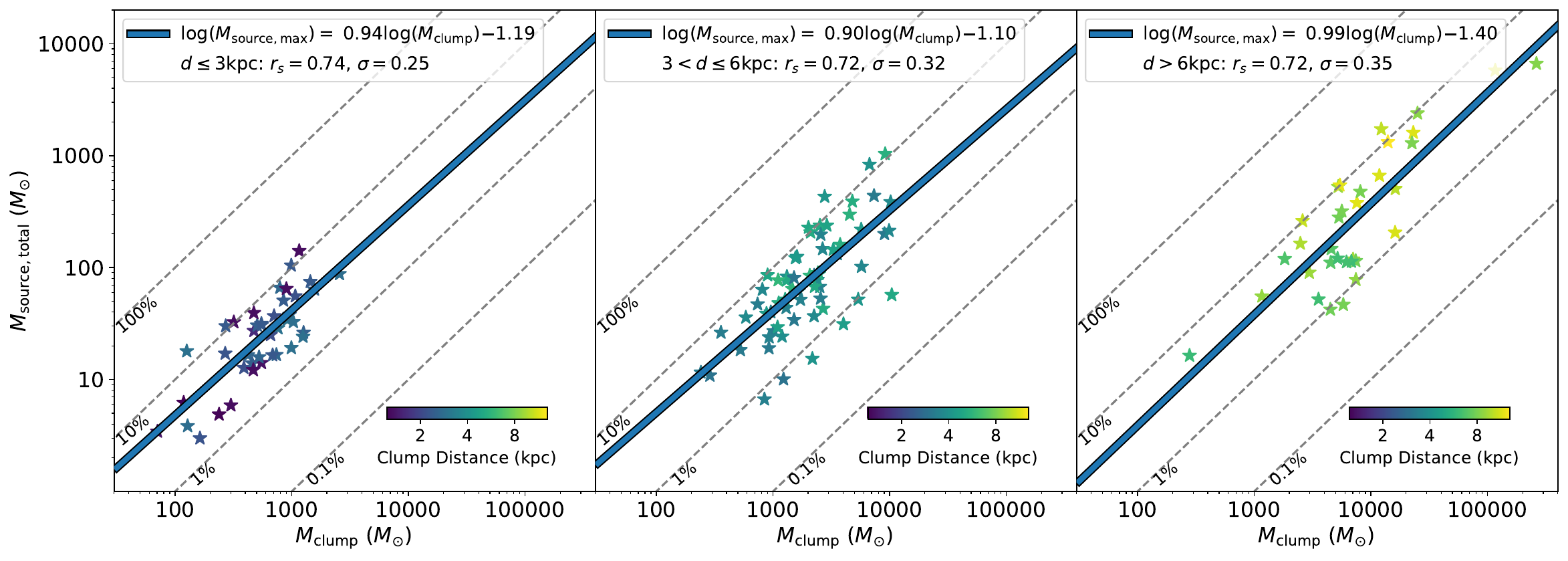}
    \caption{The correlation between the QUARKS clump mass $M_{\rm clump}$ and the total ACA source mass $M_{\rm source,total}$ in different distance bins of $d\leq3$\,kpc (left), $3<d\leq6$\,kpc (middle), and $d>6$\,kpc (right). The dashed lines label the case where $M_{\rm source,max}$/$M_{\rm source,total}$ equals to 0.1, 1, 10, and 100 percent of $M_{\rm clump}$. Linear regression is performed to fit the data in logarithmic space, and the fitting result is shown with a blue solid line. The clump distances are coded in the colors of the stars. \label{fig:DistanceEffect}}
\end{figure*}

\clearpage
\bibliographystyle{raa}
\bibliography{finale}

\end{CJK*}
\end{document}